\documentclass[aps,pra,reprint,showpacs,nolongbibliography]{revtex4-1}  

\usepackage{graphicx}
\usepackage{multirow}
\usepackage{amsmath,amsfonts,amssymb}
\usepackage{bm}
\usepackage{dcolumn}
\usepackage{graphicx}
\usepackage{nicefrac}
\usepackage{tablefootnote}
\usepackage{xcolor}

\begin{document}

\title{Pr$^{10+}$ as a candidate for a high-accuracy optical clock for tests of fundamental physics}

\author{S. G. Porsev$^{1}$}
\author{C. Cheung$^{1}$}
\author{M. S. Safronova$^{1}$}
\author{H. Bekker$^2$} 
\author{N.-H. Rehbehn$^3$} 
\author{J. R. Crespo L\'opez-Urrutia$^3$}
\author{S. M. Brewer$^4$}

\affiliation{
$^1$Department of Physics and Astronomy, University of Delaware, Newark, Delaware 19716, USA \\
$^2$Helmholtz-Institut, GSI Helmholtzzentrum f{\"u}r Schwerionenforschung, 55128 Mainz, Germany\\
$^3$Max-Planck-Institut f{\"u}r Kernphysik, 69117 Heidelberg, Germany \\
$^4$Department of Physics, Colorado State University, Fort Collins, Colorado 80523, USA}

\date{\today}

\begin{abstract}
We propose In-like Pr$^{10+}$ as a candidate for the development of a high-accuracy optical clock with high sensitivity to a time variation of the fine-structure constant, $\dot{\alpha}/\alpha$, as well as favorable experimental systematics. We calculate its low-lying energy levels by combining the configuration interaction and the coupled cluster method, achieving uncertainties as low as $0.1\%$, and improving previous work. We benchmark these results by comparing our calculations for the $|5s^25p~{^2\!P_{1/2}}\rangle \rightarrow |5s^25p~ {^2\!P_{3/2}}\rangle$ transition in Pr$^{10+}$ with a dedicated measurement and for Pr$^{9+}$ with a recent experiment, respectively. In addition, we report calculated hyperfine-structure constants for the clock and logic states in Pr$^{10+}$.  
\end{abstract}

\maketitle
\section{Introduction}

Optical atomic clocks belong to the most promising platforms for low-energy fundamental physics tests~\cite{Safronova2018RMP}. Their frequency ratios were used for laboratory-based investigations of general relativity~\cite{Chou2010Science,Bothwell2022Nature,Zheng2022Nature}, searches for signatures of ultralight scalar dark matter~\cite{Bothwell2022Nature,BACON2021Nature}, and tests for possible violations of local Lorentz invariance and local position invariance~\cite{Sanner2019Nature,Lange2021PRL}. They also yield constraints on a possible time variation of fundamental constants, such as the proton-to-electron mass ratio ($m_{p}/m_{e}$) or the fine-structure constant ($\Dot{\alpha}/\alpha$)~\cite{Rosenband2008Science,Godun2014PRL,Huntemann2014PRL,Filzinger2023PRL}.

Over the last few decades, optical clock development has focused on trapped, singly charged ions (i.e., Al$^{+}$, Yb$^{+}$), or ensembles of neutral atoms (i.e., Sr, Yb)~\cite{Ludlow2015RMP}. More recently, clocks based on laser-accessible transitions in highly charged ions (HCIs) or the nuclear clock transition in $^{229}$Th have been proposed to enhance sensitivity to signatures of new physics~\cite{Safronova2018RMP,Kozlov2018RMP,Zhang24}. In particular, strong relativistic effects in HCIs make their optical transitions the most sensitive to $\Dot{\alpha}/\alpha$~\cite{Berengut2010PRL,berengut_electron-hole_2011,berengut2012a,Safronova2014PRL,ong2014optical,Dzuba2015} (for a review, see \cite{Kozlov2018RMP}) in atomic systems. Hyperfine transitions in heavy hydrogen-like \cite{Schiller2007PRL} and Li-like \cite{Oreshkina2017} were also proposed for such studies, since they are extreme examples of relativistic effects on the binding energy of the electron and are accessible by lasers. The high charge state also largely suppresses detrimental systematic frequency shifts due to perturbations such as ac Stark shifts from blackbody radiation (BBR), as required for the most stringent clock comparisons~\cite{BACON2021Nature,Filzinger2023PRL}.

In this work, we propose a narrow linewidth transition in Pr$^{10+}$ to develop a high-accuracy optical clock for fundamental physics tests and predict its value. Both Pr$^{10+}$ and Pr$^{9+}$ possess $\alpha$-sensitive ``clock" transitions, as well as laser-accessible ``logic" transitions that are well suited for state readout using the quantum-logic spectroscopy (QLS) technique~\cite{Schmidt2005Science}. We consider the experimental feasibility of a quantum-logic clock based on Pr$^{10+}$ and calculate crucial properties for this purpose, such as hyperfine coefficients $A$ and $B$. We also estimate the expected leading systematic uncertainties and frequency instability, and evaluate the possibility of improving current bounds on $\Dot{\alpha}/\alpha$ by comparing the frequency of our proposed clock with the one based on the electric-octupole ($E3$) transition in Yb$^{+}$~\cite{Godun2014PRL,Huntemann2016PRL}.

To validate our predictions, we also measured the Pr$^{10+}$ $|5s^25p~{^2\!P_{1/2}}\rangle \rightarrow |5s^25p~ {^2\!P_{3/2}}\rangle$ M1 transition with an electron beam ion trap (EBIT) and found excellent agreement. Our present Pr$^{10+}$ theory results also match well with previous calculations but achieve higher precision than those, which is critical for upcoming experimental searches for the ultra-narrow clock transition~\cite{Safronova2014PRL}. Moreover, we also calculate the level energies for the Pr$^{9+}$ ion energies, which also agree with uncertainties from an earlier measurement~\cite{Bekker2019}.  

These theoretical benchmarks are important for other proposed clocks based on Cf$^{15+}$ and Cf$^{17+}$~\cite{Cf}. The uncertainty of predictions for clock transitions in such open-shell ions is large because of the effect of triple excitations in the coupled cluster part of the computation. We now improve the treatment of these corrections by including core-triple excitations. The agreement of the Pr$^{9+}$ and Pr$^{10+}$ experiments with the present calculations strengthens our confidence in this method for future predictions of Cf.
\section{Atomic Structure Calculations}
We use a hybrid approach that combines the configuration interaction (CI) and coupled-cluster (all-order) method. The CI wave function is obtained as a linear combination of all distinct states of a given angular momentum $J$ and parity:
\begin{equation}
\Psi_J = \sum_i c_i\Phi_i.
\end{equation}

Low-lying energies and wave functions are determined by diagonalizing the effective Hamiltonian
\begin{equation}
H^\mathrm{eff}=H_1+H_2,
\end{equation}
where $H_1$ and $H_2$ represent the one- and two-electron parts of the Hamiltonian, respectively. The CI+all-order approach allows one to incorporate core excitations in the CI method by including dominant core-core and core-valence correlation corrections into the effective Hamiltonian for all orders. We thus include in the one-electron part $H_1$ a correlation potential $\Sigma_1$, which accounts for its core-valence correlations:
\begin{equation}
H_1\rightarrow H_1+\Sigma_1,
\end{equation}
while also including in the two-electron part $H_2$ the respective core-valence interactions:
\begin{equation}
H_2\rightarrow H_2+\Sigma_2.
\end{equation}
A detailed description of the CI+all-order method and all formulas is given in Ref.~\cite{Safronova2009}.
We use the parallel CI package developed in Ref.~\cite{sym2021} for all calculations.

We begin our study with Sn-like Pr$^{9+}$, where experimental data are available~\cite{Bekker2019}. To evaluate the uncertainties of our results, we performed several calculations to separate the contributions of higher-order correlations, higher partial waves, triple excitations, and quantum-electrodynamical (QED) contributions. We then used the same approach to study Pr$^{10+}$, for which no experimental data were available before this work.

We consider both Pr$^{9+}$ and Pr$^{10+}$ ions as atomic systems with $[1s^2\dots 4d^{10}]$ closed core shells and open $5s$, $5p$, $5d$, $6s$, $6p$ and $4f$ valence shells. We used a complete set of Dirac-Hartree-Fock wave functions on a radial grid generated using $B$-splines constrained to a spherical cavity of radius $R=20$ a.u. The basis set is constructed using 40 splines of order 7. The Breit interaction is included in the stage of constructing the basis set. 
We treat Pr$^{9+}$ as a 4-valent system and start with all possible single and double excitations to any orbital up to $22spdfg$ from the $5s^2 5p^2$ and $5s^2\, 5p 4f$ even configurations (in the following, we leave out the $5s^2$ from the configuration designations for brevity). Here, $22spdfg$ indicates that excitations to all orbitals up to $n=22$ are included for the $spdfg$ partial waves.

In Table~\ref{pr9_basis}, we list the energies obtained from increasing sets of configurations (from $22spdfg$ to $30spdfg$)
and show their convergence. The energy differences between $22spdfg$ and $26spdfg$ do not exceed $74$ cm$^{-1}$,
while the largest difference between $26spdfg$ and $30spdfg$ is only about 3 cm$^{-1}$.
This clear convergence of CI allows us to use the set of configurations $22spdfg$ for subsequent calculations, which is sufficiently complete but does not make the calculations too time-consuming.

The contributions to the excitation energies of Pr$^{9+}$ from other correlation effects are listed in Table~\ref{pr9_table}.
The results are compared with the experimental data from~\cite{Bekker2019}.
We added corrections due to three-electron excitations from the main configurations and allowed single and double excitations
only from the $5s^2 4f^2$ configuration, since including additional main configurations had a negligible effect.
The calculated corrections, listed in columns ``triples'' and ``$4f^2$'' of Table~\ref{pr9_table}, are relatively small and partially cancel each other.
The energy differences between $22spdfg$ and $30spdfg$ are given in the column labeled ``(23-30)$spdfg$.''
We then performed calculations for $30spdfgh$ and $30spdfghi$, including excitations from the main configurations to
the $(6-30)h$ and $(7-30)i$ orbitals, and list the respective corrections in the columns labeled ``(6-30)$h$'' and ``(7-30)$i$.''
\begin{table}[tbp]
\caption{\label{pr9_basis} Energy corrections for Pr$^{9+}$ calculated with increasing CI space size (all energies are given
in cm$^{-1}$) for the ${\it n}\,spdfg$ sets of configurations, where {\it n} is the largest principal quantum number
for each included partial wave.}
\begin{ruledtabular}
\begin{tabular}{rcccc}
\multicolumn{1}{c}{Term} &
\multicolumn{1}{c}{$22spdfg$} & \multicolumn{1}{c}{$26spdfg$} & \multicolumn{1}{c}{$30spdfg$} \\
\hline \\ [-0.6pc]
$5p^2  \,\, ^3\!P_0$  &    0    &   0   &   0   \\
$5p 4f \,\, ^3\!G_3$  &  22815  &  -66  &  -2   \\
$5p 4f \,\, ^3\!F_2$  &  24915  &  -66  &  -3   \\
$5p 4f \,\, ^3\!D_3$  &  27956  &  -70  &  -2   \\
$5p^2  \,\, ^3\!P_1$  &  28462  &    7  &   0   \\
$5p 4f \,\, ^3\!G_4$  &  30288  &  -74  &  -3   \\
$5p^2  \,\, ^3\!D_2$  &  36453  &  -15  &  -1   \\
$5p 4f \,\, ^3\!F_3$  &  56151  &  -63  &  -2   \\
$5p 4f \,\, ^3\!F_4$  &  59869  &  -68  &  -2   \\
$5p^2  \,\, ^3\!F_2$  &  62748  &  -54  &  -2   \\
$5p 4f \,\, ^3\!G_5$  &  65030  &  -65  &  -2   \\
$5p 4f \,\, ^1\!F_3$  &  64843  &  -73  &  -2   \\
$5p^2  \,\, ^3\!P_2$  &  67777  &  -45  &  -2   \\
$5p 4f \,\, ^3\!D_1$  &  68481  &  -71  &  -3
  \end{tabular}
\end{ruledtabular}
\end{table}
\begin{table*}[tbp]
\caption{\label{pr9_table} Contributions to the excitation energies of Pr$^{9+}$ (in cm$^{-1}$) from different correlation effects (all energies are given in cm$^{-1}$).
The final excitation energies (given in the column labeled ``Final'') are calculated by adding the corrections (see the main text) listed in columns 3-11 to the base result $22spdfg$. The experimental values of~\cite{Bekker2019} are shown in the column labeled ``Exp.\cite{Bekker2019}''. The differences between the final theoretical and experimental results are given (in cm$^{-1}$) in the column ``Diff.''.}
\begin{ruledtabular}
\begin{tabular}{rccccccccccccc}
\multicolumn{1}{c}{Term\cite{Bekker2019}}&
\multicolumn{1}{c}{$22spdfg$}&
\multicolumn{1}{c}{triples$^a$}&
\multicolumn{1}{c}{$4f^2$}&
\multicolumn{1}{c}{(23-30)$spdfg$}&
\multicolumn{1}{c}{(6-30)$h$}&
\multicolumn{1}{c}{(7-30)$i$}&
\multicolumn{1}{c}{NLTr}&
\multicolumn{1}{c}{QED}&
\multicolumn{1}{c}{TEI}&
\multicolumn{1}{c}{Extrap.}&
\multicolumn{1}{c}{Final} &
\multicolumn{1}{c}{Exp.\cite{Bekker2019}}&
\multicolumn{1}{c}{Diff.}\\
\hline \\ [-0.6pc]
$5p 4f \,\, ^3\!G_3$       &  22815 &  26  &  -4  &  -69  & -217  &  -41  &  893 &  -199 & -107 & -1065 & 22032  & 22101 &  -69 \\
$5p 4f \,\, ^3\!F_2$       &  24915 &  18  &  -4  &  -68  & -191  &  -35  &  631 &  -152 &  219 &  -885 & 24448  & 24494 &  -46 \\
$5p 4f \,\, ^3\!D_3$       &  27956 &  21  &  -6  &  -72  & -222  &  -42  &  840 &  -123 &   18 & -1058 & 27310  & 27287 &   23 \\
$5p^2  \,\, ^3\!P_1$       &  28462 &  18  &   8  &    7  &   29  &   13  &  -69 &    61 &   37 &     7 & 28573  & 28561 &   12 \\
$5p 4f \,\, ^3\!G_4$       &  30288 &  20  &  -8  &  -77  & -278  &  -59  &  828 &  -124 & -258 & -1066 & 29266  & 29231 &   35 \\
$5p^2  \,\, ^3\!D_2$       &  36453 &   4  &  -2  &  -15  &  -56  &  -11  &  181 &     4 &  113 &  -234 & 36436  & 36407 &   29 \\
$5p 4f \,\, ^3\!F_3$       &  56151 &  23  &  -7  &  -66  & -194  &  -36  &  832 &  -144 &  178 & -1057 & 55681  & 55662 &   19 \\
$5p 4f \,\, ^3\!F_4$       &  59869 &  22  &  -8  &  -70  & -210  &  -40  &  794 &  -104 &   16 & -1054 & 59215  & 59185 &   30 \\
$5p^2  \,\, ^3\!F_2^b$     &  62748 &  12  &  -6  &  -56  & -174  &  -30  &  434 &   -73 &   -2 &  -753 & 62101  & 62182 &  -81 \\
$5p 4f \,\, ^3\!G_5$       &  65030 &  27  &  -3  &  -67  & -239  &  -46  &  737 &   -69 & -483 & -1045 & 63842  & 63924 &  -82 \\
$5p 4f \,\, ^1\!F_3^c$     &  64843 &  17  &  -7  &  -76  & -238  &  -43  &  720 &   -81 & -190 & -1047 & 63898  & 63964 &  -66 \\
$5p^2  \,\, ^3\!P_2^{d}$  &  67777 &  7   &  -7  &  -47  & -142  &  -22  &  347 &   -24 & -100 &  -614 & 67175  & 67291 & -116 \\
$5p 4f \,\, ^3\!D_1$       &  68481 &  16  &  -4  &  -74  & -267  &  -48  &  694 &  -144 & -542 & -1049 & 67063  & 67309 & -246
\end{tabular}
\end{ruledtabular}
\begin{flushleft}
\footnotesize{
$^a$These terms were calculated for $12spdfg$. \\
$^b$This term must have a typo in Ref.~\cite{Bekker2019} as the $5p^2$ configuration cannot have $L=3$; we determined it as $5p4f \,\, ^3\!D_2$. \\
$^c$In this work, we determined this term as $^3\!D_3$. \\
$^d$We determined this term as $^3\!D_2$ with the following weights of the two main configurations: 57\% $5p4f$ and 38\% $5p^2$.
}
\end{flushleft}
\end{table*}

Using the expressions for the cluster amplitudes derived in Ref.~\cite{PorDer06}, we included the nonlinear (NL) terms and triple excitations into the formalism of the CI+all-order method developed in Ref.~\cite{Safronova2009}. The total corrections for these terms
are presented in the column labeled ``NLTr''.
Following Refs.~\cite{QED} and~\cite{TEI}, respectively, we included QED corrections and three-electron interaction (TEI) corrections. They are given in the columns ``QED'' and ``TEI''.

Additionally, we explored the contributions of higher partial waves in the all-order expansion. Partial waves with $l_\mathrm{max}=6$ are included in every summation in all-order terms for each of the calculations. To explore the role of higher partial waves,
we also performed a complete all-order calculation with $l_\mathrm{max}=7$ and $l_\mathrm{max}=8$ and found
that the inclusion of partial waves with $l>6$ is very important for an accurate description of the $5p 4f$ states. We extrapolate the contribution $l>6$ following the method described in Ref.~\cite{SafDzuFla14b}. We estimate the effect of higher partial waves on Pr$^{9+}$ as the difference between the results obtained for $l_\mathrm{max}=8$ and $l_\mathrm{max}=6$ to which we added
the difference between the results obtained for $l_\mathrm{max}=8$ and $l_\mathrm{max}=7$. This contribution is labeled ``Extrap.''
in Table~\ref{pr9_table}.

The comparison of theoretical and experimental excitation energies for Pr$^{9+}$
(the last column, ``Diff.'', in Table~\ref{pr9_table}) shows that the difference does not exceed 70 cm$^{-1}$ for most states belonging to the $5p4f$ configuration.
For the difference between the fine-structure states $5p^2  \,\, ^3\!P_0$ and $5p^2  \,\, ^3\!P_1$, the theory differs from the experiment by only 12 cm$^{-1}$.
This remarkable agreement shows the predictive power of this approach for Pr$^{10+}$. Although the effect of triple excitations is even larger in Cf ions, we expect that the inclusion of core triple excitations will allow us to improve the accuracy of the predicted clock wavelength from that of Ref.~\cite{Cf}.
\subsection{Pr$^{10+}$ Energies}
\begin{figure*}[]
\centering
\includegraphics[width=1.4\columnwidth, angle=0]{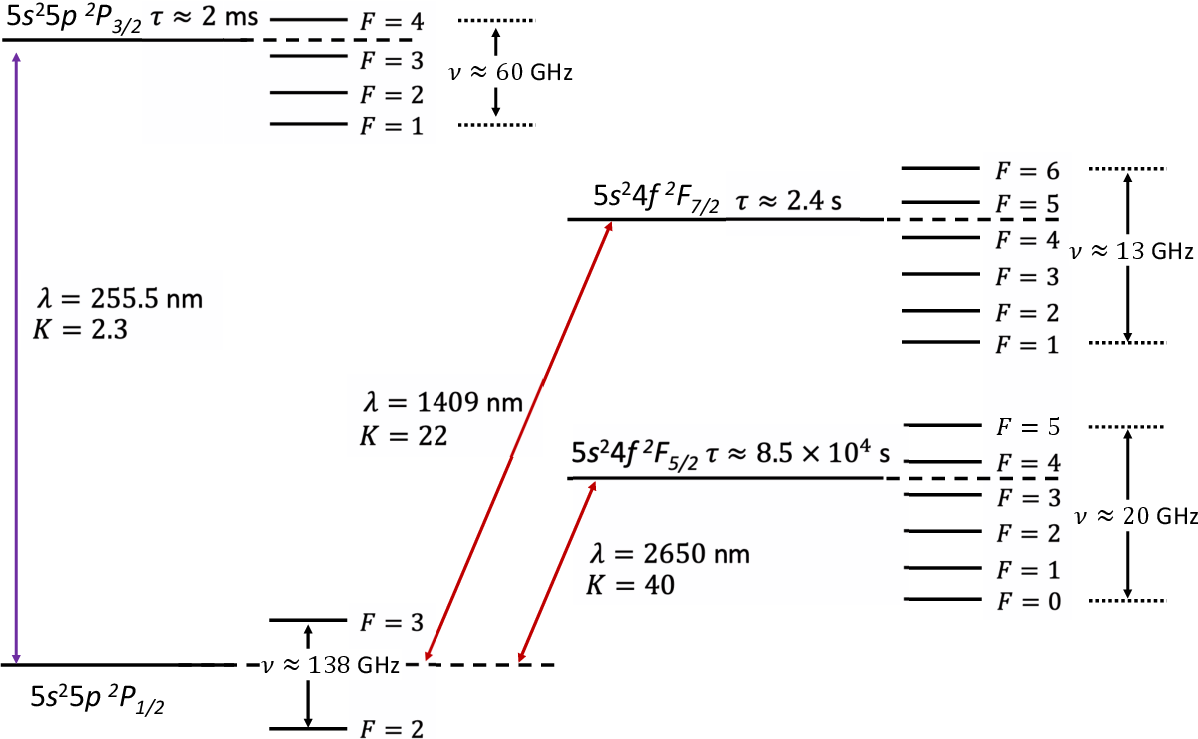}
\caption{Energy levels, radiative lifetimes, and hyperfine-structure of the lowest states in $^{141}$Pr$^{10+}$,
         with nuclear spin, $I = 5/2$.}
\label{fig: PrXI-levels}
\end{figure*}

Our treatment of Pr$^{10+}$ as a trivalent system starts with all possible single and double excitations to any orbital up to $26spdfg$
from the $5s^2 5p$ and $5s^2 4f$ odd main configurations. A simplified energy-level diagram highlighting the transitions of interest is shown in Fig.~\ref{fig: PrXI-levels}.
In Table~\ref{pr10_table}, we list the contributions to the Pr$^{10+}$ energies from different correlation effects.  
Excitations from additional main configurations (AMC) $5p^2 4f$, $5p^3$, $5s 5p 5d$, $5s 6s 4f$, and $5s5p6s$ give a small contribution to the excitation energies. They are presented in a column
labeled ``AMC'' in Table~\ref{pr10_table}. By performing the calculation for $30spdfg$, we determined contributions from configurations
obtained by excitations to 27-30$spdfg$ shells (labeled as ``(27-30)$spdfg$'').
All other contributions to the excitation energies listed in the table are analogous to those for Pr$^{9+}$ described above.

Based on the results for the excitation energies of Pr$^{9+}$ and considering that the calculation for Pr$^{10+}$ was performed similarly, producing comparable precision, we assign uncertainties to the energies of the excited states. The final values with their uncertainties are given in Table~\ref{pr10_table} in the column ``Final''.
\begin{table*}[tbp]
\caption{\label{pr10_table} Contributions to the excitation energies of Pr$^{10+}$ (in cm$^{-1}$)
from different correlation effects. The final excitation energies (given in the column labeled ``Final'') are calculated by adding the
corrections listed in columns 3??11 to the base result $26spdfg$. The theoretical
results of~\cite{Safronova2014PRL} are presented for comparison. The uncertainties are given in parentheses.}
\begin{ruledtabular}
\begin{tabular}{rcccccccccccc}
  \multicolumn{1}{c}{Term} & \multicolumn{1}{c}{$26spdfg$} & \multicolumn{1}{c}{triples$^a$} & \multicolumn{1}{c}{AMC$^b$}
& \multicolumn{1}{c}{(27-30)$spdfg$} & \multicolumn{1}{c}{(6-30)$h$} & \multicolumn{1}{c}{(7-30)$i$} & \multicolumn{1}{c}{NLTr}
& \multicolumn{1}{c}{QED} & \multicolumn{1}{c}{TEI} & \multicolumn{1}{c}{Extrap.} & \multicolumn{1}{c}{Final}
& \multicolumn{1}{c}{Ref.~\cite{Safronova2014PRL}} \\
\hline \\ [-0.6pc]
$4f_{5/2}$       &   4372 &   2  &   2  &  -2   & -178  &  -43  &  845 &  -98  &  -37 & -1090 & $3773(70)$  & $3702(200)$ \\
$4f_{7/2}$       &   7734 &   2  &   2  &  -2   & -180  &  -43  &  810 &  -97  &  -43 & -1083 & $7099(70)$  & $7031(200)$ \\
$5p_{3/2}$       &  39139 &   0  &   0  &   0   &    5  &    2  &  -69 &   50  &    7 &     7 & $39140(15)$ & $39141(40)$ \\
\end{tabular}
\end{ruledtabular}
\begin{flushleft}
\footnotesize{
$^a$These terms were calculated for $12spdfg$, \\
$^b$Contribution from electron excitations from additional main configurations $5p^2 4f$, $5p^3$, $5s5p5d$, $5s6s4f$, and $5s5p6s$.
}
\end{flushleft}
\end{table*}
\subsection{Hyperfine-structure constants}
We also calculated the magnetic-dipole and electric-quadrupole hyperfine-structure (hfs) constants $A$ and $B$ of the four lowest-lying states. For $^{141}$Pr$^{10+}$, the nuclear spin is $I=5/2$ and the nuclear magnetic dipole and electric quadrupole moments are $\mu/\mu_N = 4.2754(5)$ (where $\mu_N$ is the nuclear magneton) and $Q= -0.077(7) \, {\rm b}$, respectively~\cite{Sto05}.

The results obtained for the $[26spdfg]$ set of configurations are summarized in Table~\ref{hfs}. In the third column, we present the results for the $g$ factors of the states obtained in the CI+all-order approximation. 
In the fourth, we give the results obtained at the CI+MBPT stage
when the core-valence correlations are included in the second order of the perturbation theory. CI+all-order results that include higher-order core-valence correlations are displayed in the column labeled ``CI+All.'' 
Finally, we included random phase approximation (RPA) corrections; the results are listed in the column labeled ``CI+All+RPA.''
We use these results to arrive at the final values listed in column ``Final.''

Based on the difference between the CI+all-order and CI+MBPT results, we determine the uncertainties for the constants $A$. For the constants $B$,
the difference between the CI+all-order and CI+MBPT is small, while the RPA corrections change the values of $B(4f_{5/2,7/2})$ by more than 20\%.
Here, we do not take into account corrections to the hyperfine operator beyond RPA, such as the core Brueckner, structural radiation, two-particle corrections, and normalization (see Ref.~\cite{DzuKozPor98} for more details), which can affect the values of the constants $B$ noticeably. For this reason, we consider their final values as estimates
sufficiently accurate for our present purpose of analyzing experimental measurement and assessing proposed clock accuracy. 
\begin{table}[tbp]
\caption{Calculated $g$ factor and hfs constants $A$ and $B$ (in MHz).
The CI+MBPT and CI+all-order values are listed in the columns labeled ``CI+MBPT'' and ``CI+All,'' respectively. The values including the RPA corrections, are listed in the column labeled ``CI+All+RPA.'' The final values are given in the column labeled ``Final.'' Uncertainties are given in parentheses.}
\label{hfs}%
\begin{ruledtabular}
\begin{tabular}{lcccccc}
       &             &  $g$     & CI+MBPT & CI+All & CI+All+RPA &    Final   \\    
\hline \\ [-0.5pc]
  $A$  &  $5p_{1/2}$ &  0.666   &  46318  & 42722  &  46070     & $46100(3600)$   \\[0.1pc]
       &  $4f_{5/2}$ &  0.856   &   1690  &  1277  &   1313     &  $1300(400)$   \\[0.1pc]
       &  $4f_{7/2}$ &  1.142   &    773  &   965  &    675     &   $670(190)$   \\[0.1pc]
       &  $5p_{3/2}$ &  1.333   &   7153  &  5736  &   6596     &  $6600(1400)$   \\[0.3pc]
\hline \\ [-0.5pc]
  $B$  &  $4f_{5/2}$ &          &    -69  &   -67  &    -84     &     -85         \\[0.1pc]
       &  $4f_{7/2}$ &          &    -79  &   -77  &    -98     &    -100         \\[0.1pc]
       &  $5p_{3/2}$ &          &   -571  &  -528  &   -586     &    -585
\end{tabular}
\end{ruledtabular}
\end{table}

\subsection{Electric-quadrupole moments and polarizabilities}
We also calculate the electric quadrupole moments and polarizabilities of the ground and excited logic and clock states shown in Fig.~\ref{fig: PrXI-levels}.  
The quadrupole moment $\Theta$ of an atomic state  is given by
\begin{eqnarray}
\Theta &=& 2\, \langle J, M=J |Q_0| J, M=J \rangle \nonumber \\
       &=& 2\, \sqrt{\frac{J(2J-1)}{(2J+3)(J+1)(2J+1)}} \langle J ||Q|| J \rangle ,
\end{eqnarray}
where $ \langle J ||Q||  J \rangle$ is the reduced matrix element of the electric quadrupole operator.

The static polarizability $\alpha_{JM}$ of a $|JM\rangle$ state 
can be expressed as a sum over unperturbed intermediate states:
\begin{equation}
\alpha_{JM} = 2 \sum_{n} \frac{|\langle JM|D_z|J_n M\rangle|^2}{E_n-E_J},
\label{alpha}
\end{equation}
where $\bf D$ is an electric dipole moment operator, and $E_J$ and $E_n$ are the energy
of the state $|JM\rangle$ and an intermediate state $|J_n M\rangle$, respectively.

The results are summarized in Table~\ref{E2_pol}. The values of the static polarizabilities calculated at $M=1/2$ are displayed in the second row. The third row shows the differential polarizabilities between the excited and ground states. The electric quadrupole moments in $ea_B^2$, where $a_B$ is the Bohr radius, are given in the fourth row. 
\begin{table}[tbp]
\caption{Electric-quadrupole moments $\Theta$ (in $e a_B^2$) and polarizabilities (in a.~u.) of the ground and excited logic and clock states.}
\label{E2_pol}%
\begin{ruledtabular}
\begin{tabular}{lccc}
Term       &  $\alpha_{JM=1/2}$ & $\Delta \alpha$ & $\Theta$ \\    
\hline \\ [-0.5pc]
$5p_{1/2}$ &        1.72        &                 &         \\[0.1pc]
$4f_{5/2}$ &        1.77        &    0.05         &  -0.34  \\[0.1pc]
$4f_{7/2}$ &        1.77        &    0.05         &  -0.41  \\[0.1pc]
$5p_{3/2}$ &        1.65        &   -0.07         &  -0.87
\end{tabular}
\end{ruledtabular}
\end{table}

\section{P\lowercase{r}$^{10+}$ measurements}
To support this work, spectroscopic measurements of the $|5s^25p~{^{2}P_{1/2}}\rangle \rightarrow |5s^25p~ {^{2}P_{3/2}}\rangle$ transitions of Pr$^{10+}$ were made (as for Pr$^{9+}$) with the Heidelberg EBIT (HD-EBIT). It produces HCI by electron impact ionization with a 3.8~mA electron beam accelerated by a potential of 172~V and focused in an 8-T magnetic field~\cite{crespolopez-urrutiaProgressHeidelbergEBIT2004}. Praseodymium ions were prepared by injecting an organometallic compound (CAS number 15492-48-5) into the EBIT from a cell. By heating it to 120~$^{\circ}$C, a pressure of $6\cdot10^{-8}$~mbar was established in the second stage of a differential pumping system, generating a tenuous molecular beam that crosses the electron beam at the center of the trap and is dissociated by it, freeing Pr atoms that are immediately ionized by the beam. The produced ions are then trapped, the trap depth being made sufficiently shallow by applying potentials to the drift tubes such that light ions of carbon, oxygen, and hydrogen (resulting from the dissociation of the compound) evaporate from the EBIT, leaving a rather pure ensemble of the much heavier ions of Pr, as pioneered in \cite{Levine1988,Penetrante1991}. By tuning the depth of the trap and regularly ejecting the contents of the trap, which tends to accumulate contaminant ions from the cathode materials Ba and W over time scales of minutes, we ensured that Pr$^{10+}$ was the dominant species in the trap at nearly all times.

\begin{figure}[ht]
    \centering
    \includegraphics[width=1.0\columnwidth]{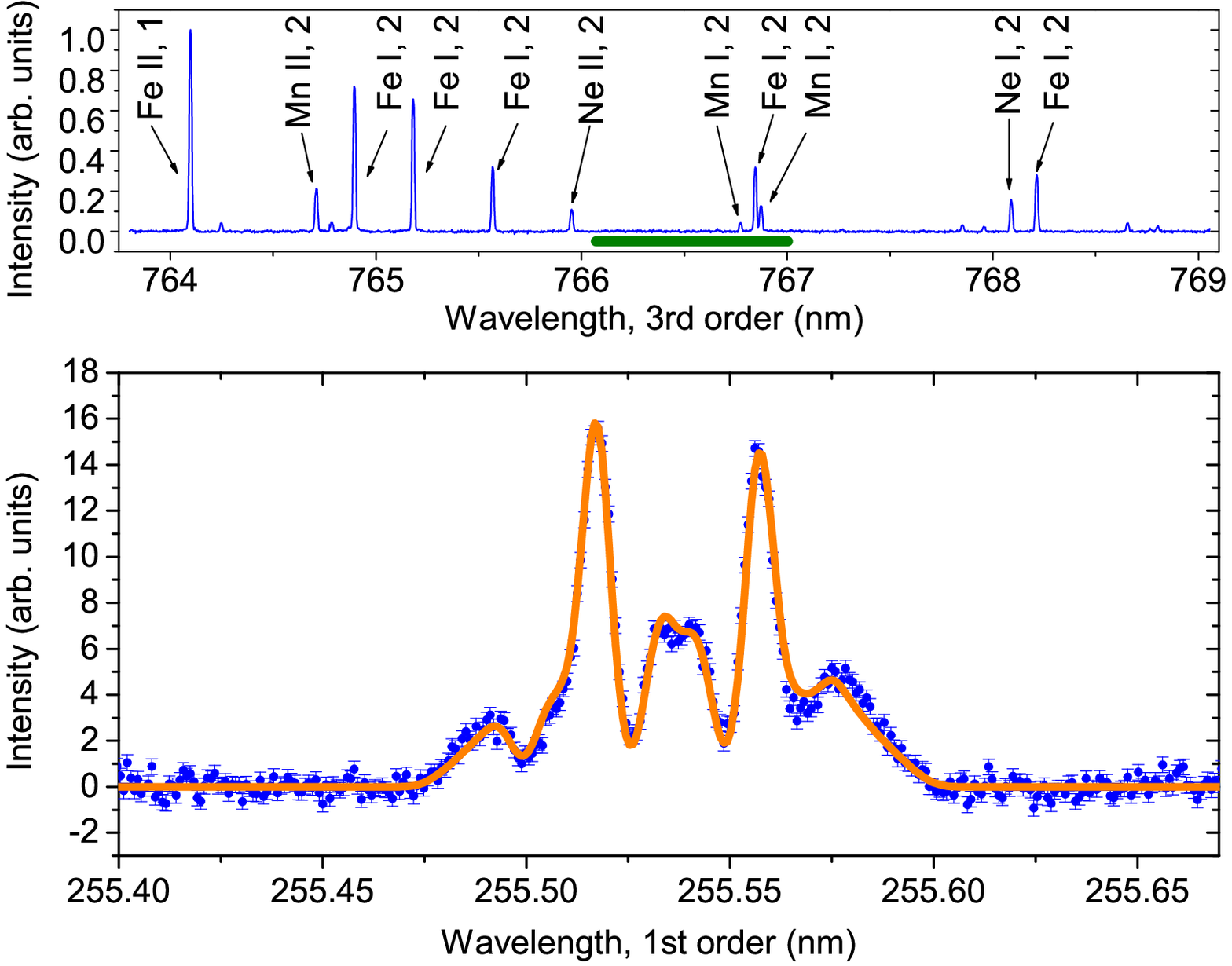}
    \caption{Top: Spectrum of the Fe-Mn-Cu-Ne calibration lamp. The identified species are labeled, with the second number indicating the diffraction order in which each line appears. The green horizontal line indicates the range displayed in the bottom graph. Bottom: Measured spectrum of the $^2\!P_{1/2}$ - $^2\!P_{3/2}$ Pr$^{10+}$ transition from 31 hours of exposures (blue) and fit (orange) used to determine the transition energy.}
    \label{fig: PrXI-spectrum}
\end{figure}

Continuous electron impact excitation of Pr$^{10+}$ and subsequent decays generated fluorescence light which was focused on the entrance slit of a grating spectrometer using four lenses and three mirrors. The 2-m focal length spectrometer was fitted with a 3600~lines/mm grating. By operating in third diffraction order, a linear dispersion of 0.9~pm/pixel at 255~nm was achieved~\cite{bekkerLaboratoryPrecisionMeasurements2018} in the focal plane where the Peltier-cooled CCD camera is installed. A total of 31 Pr$^{10+}$ spectra were recorded for 1 hour each. These were interspersed with 8 spectra obtained without injection of Pr for background subtraction. Calibration spectra were obtained before each Pr$^{10+}$ measurement from a Fe-Mn-Cu-Ne hollow cathode lamp that illuminates a movable diffuse reflector at an intermediate focus of the imaging system. The integrated total Pr$^{10+}$ spectrum and a sample calibration spectrum are shown in Fig.~\ref{fig: PrXI-spectrum}. Although Pr$^{10+}$ can be easily produced in compact room-temperature EBITs, the strong magnetic field (up to 8~T, in this case, 5.77~T) of the HD-EBIT yields large Zeeman splittings of the observed transitions, which can be resolved despite Doppler broadening caused by temperatures of the trapped ions here on the order of $10^5$\,K. This allows for unambiguous identification of lines or, conversely, the determination of $g$ factors and hfs constants.

In the present case, the hyperfine interaction is of similar magnitude as the Zeeman one, so the Breit-Rabi formalism has to be applied. We construct the Hamiltonian on the $F, m_F$ basis relying on the $g$ factors and the hfs $A$ constants from Table~\ref{hfs} and determine the eigenvalues and eigenvectors. These are used to model the line shape, which is fit to the data as in Ref.~\cite{tremblayAbsorptionProfilesAlkalimetal1990}. The result for the transition energy is 39122.234(13)~cm$^{-1}$, where the 1-$\sigma$ uncertainty is mainly due to calibration uncertainties. The only other free fit parameters were a Gaussian width common to all Zeeman components and the ratio between the grating efficiency for $s$- and $p$-polarization due to the Zeeman effect.
This measured transition energy is in good agreement with the theoretical prediction of 39140(15) cm$^{-1}$ listed in Table~\ref{pr10_table}. 

\section{Optical Clock Considerations}
Here, we consider the experimental feasibility of developing an optical clock based on Pr$^{10+}$.  Both clock transitions, ${|^2P_{1/2}\rangle \rightarrow |^2F_{7/2}\rangle}$ and ${|^2P_{1/2}\rangle \rightarrow |^2F_{5/2}\rangle}$, offer increased sensitivity to possible time-variation of the fine-structure constant ($\dot{\alpha}/\alpha$) and better systematics than current state-of-the-art ion clocks~\cite{Brewer2019PRL,Filzinger2023PRL}.  For technical considerations related to laser performance and frequency instability, we will focus on the ${|^2P_{1/2}\rangle \rightarrow |^2F_{7/2}\rangle}$ transition, which is predicted to have a wavelength of ${\lambda = 1409}$~nm. Due to the relatively low ionization energy of $\approx 162$~eV, Pr$^{10+}$ has also recently been produced in a compact EBIT using a laser ablation loading scheme~\cite{BanducciInPrep}. The high ion temperature inside an EBIT precludes clock operation inside it. To this end, sympathetic cooling has to be performed in a separate radiofrequency (rf) trap \cite{SchmogerScience2015}, to which the HCI has to be transferred, in this case a single Pr$^{10+}$ ion. The lack of a laser-accessible cooling and readout transition requires co-trapping of this ion with Be$^{+}$ to cool it to a temperature of $\approx 1$~mK, where QLS will be used for state readout~\cite{Schmidt2005Science}. When the ions are transferred to the rf trap, their charge state can be confirmed by measuring the motional frequencies of the Be$^{+}$/Pr$^{10+}$ ion crystal.  Depending on the trapping conditions of the EBIT, the transferred Pr$^{10+}$ ion can be in the ground electronic state or in one of the excited logic or clock states.  Based on a collisional-radiative model of the EBIT plasma, a few seconds after ion transfer to the rf trap it will be found in the ground electronic state roughly 50~\% of the time. Depending on which state the ion is in, clock operation can proceed, or the ion can be released, and another can be loaded. This procedure can be repeated until the ion is found to be in the $^2\!P_{1/2}$ state.  Once in the $^2\!P_{1/2}$ state, a particular $|F,m_F\rangle$ sublevel can be prepared using an optical pumping scheme. 

For reference, following sympathetic cooling \cite{SchmogerScience2015} and QLS \cite{Micke2020Nature}, clock operation with an accuracy of $\Delta\nu/\nu = 2.2 \times 10^{-17}$ was recently demonstrated in B-like Ar$^{13+}$~\cite{King2022Nature}. In that work, the main contribution to systematic uncertainty was the time-dilation shift from a technical excess micromotion (EMM) of the ion in the trap~\cite{King2022Nature}.  This type of EMM is typically due to trap imperfections introduced during the manufacturing process and can be reduced to a level of $10^{-19}$ by using a trap with low residual rf fields~\cite{Keller2015JAP,Keller2019PRA}.  The time dilation shift due to the secular ion motion can be reduced to a low $10^{-18}$ or below by operating the clock near the motional ground state~\cite{Chen2017PRL,Brewer2019PRL}. This can be accomplished by the use of sub-Doppler cooling (ground-state cooling) before clock interrogation, if the motional heating rates in the trap are low enough. If heating rates become problematic, periodic cooling during clock interrogation can be used to mitigate this heating.  Due to the low differential polarizability of the clock transition, the Stark shifts associated with this cooling should be negligible.

The presence of blackbody radiation (BBR) at the location of the clock ion will cause an ac Stark shift on the clock levels, leading to a frequency shift in the clock transition.  The BBR-induced frequency shift is given as ${\Delta \nu = -(1/2h) \Delta \alpha \langle E_\mathrm{BBR}^{2} \rangle}$, where $\Delta \alpha$ is the differential polarizability for a given clock transition, and ${\langle E_\mathrm{BBR}^{2} \rangle = (831.9\, \rm{V/m})^{2} (T / 300\, \rm{K})^{4}}$, where $T$ is the temperature of the BBR environment. The differential polarizabilities of the clock and logic transitions have been calculated in this work and are given in Table~\ref{E2_pol}. For the $^2\!P_{1/2} \rightarrow \, ^2\!F_{7/2}$ transition at room temperature, the BBR-induced frequency shift would be ${\Delta \nu /\nu \approx -2\times10^{-18}}$.  However, due to the high charge of Pr$^{10+}$, the trap must be operated at cryogenic temperatures, reducing its pressure to the level of $10^{-14}$ mbar to avoid ion losses by charge exchange with residual gas. At an adequate temperature of approximately 4~K, the BBR-induced frequency shift would be ${\Delta \nu /\nu \approx -6\times10^{-26}}$, which is effectively negligible.

The electric quadrupole moments calculated for the excited clock and logic states are presented in Table~\ref{E2_pol}. Due to the non-zero electric quadrupole moment in the excited clock states, an electric field gradient at the location of the ion due, e.~g., to the trapping potential, will lead to an electric-quadrupole shift on the clock transition. At usual trapping conditions~\cite{King2022Nature}, we estimate that the quadrupole shift when addressing a particular Zeeman sublevel will be on the order of $100$~mHz ($\approx 3.5\times10^{-16}$, fractionally). This shift is approximately an order of magnitude smaller than in $^{88}$Sr$^{+}$~\cite{Dube2013PRA}.  We expect that by averaging the shifts from all Zeeman sublevels, it should be possible to suppress its uncertainty to below $10^{-18}$~\cite{Dube2005PRL,Dube2013PRA}.

We also estimate the achievable clock instability. The noise-limited quantum projection instability for a clock based on a single Pr$^{10+}$ ion can be estimated by assuming a Ramsey interrogation time that is equal to the lifetime of the excited $^{2}F_{7/2}$ state~\cite{Itano1993PRA}. The clock instability is given as~\cite{Peik2005JPhysB}
\begin{equation}
    \sigma_{y}(t) = \frac{0.412}{\nu \sqrt{\tau t}},
\end{equation}
where $\nu$ is the clock frequency, $\tau$ is the lifetime of the excited state, and $t$ is the averaging time.  Under these conditions, we obtain ${\sigma_{y}(t) \approx 1.3 \times 10^{-15}/\sqrt{t/\rm{s}}}$.  This value is comparable to state-of-the-art single-ion clocks based on Al$^{+}$ and Yb$^{+}$~\cite{Brewer2019PRL, Huntemann2016PRL} and is roughly a factor of 20 lower than the instability demonstrated in the recent Ar$^{13+}$ clock~\cite{King2022Nature}.  With this frequency instability, it would be possible to perform clock comparisons at the low $10^{-18}$ level with approximately ten days of averaging time.
\section{Conclusions and prospects}
We analyzed the prospect of developing a high-accuracy optical clock based on a Pr$^{10+}$ ion and evaluated its potential systematic effects. Our calculations were contrasted with previous Pr$^{9+}$ measurements and our current measurement of the Pr$^{10+}$ the $|5s^25p~{^{2}P_{1/2}}\rangle \rightarrow |5s^25p~ {^{2}P_{3/2}}\rangle$ transition. We find excellent agreement between theory and existing measurements. 

The $\alpha$ sensitivity coefficient of Pr$^{10+}$ is large and positive, $K=22$. In particular, it is an order of magnitude larger than $K({\rm Ar}^{13+}) =1.95$~\cite{YuSahSuo23}.
We propose an initial comparison with a clock based on the $E3$ transition in Yb$^{+}$, which has a large negative $\alpha$ sensitivity ($K = -5.95$)~\cite{Godun2014PRL,Huntemann2014PRL}.  Measurements of the frequency ratio of these two clocks with an accuracy of $10^{-18}$, separated by an interval of one year, would set a limit on $\Dot{\alpha}/\alpha$ at the level of $10^{-20}/\rm{yr}$. This outcome would be at the level of recent work in Yb$^{+}$~\cite{Filzinger2023PRL} that required almost 8 years of frequency comparison data. With the development of an optical fiber network~\cite{VanArsdale2024OptLett}, it would be possible to include other clocks with different $\alpha$ sensitivities (that is, those based on Al$^{+}$, Sr, and Yb) to search for a wide range of new-physics signals~\cite{Ludlow2015RMP,Kozlov2018RMP}. Following the theoretical estimates from e.~g., Ref.~\cite{Yudin2014} and several other works, and in light of rapid advances in laser stabilization and frequency metrology techniques, we are optimistic about the tremendous potential of HCI with their extremely low polarizabilities for ultimately outperforming other clock transitions and providing a peer to the clocks based on the nuclear transition $^{229m}$Th isomeric transition, which was recently laser excited in Ref.~\cite{Tiedau2024,ElwellPRL2024} and compared with a Sr lattice clock in Ref.~\cite{Zhang24} that are under development. Frequency comparisons of clocks based on HCI and trapped $^{229m}$Th ions \cite{Peik2003EPL,Peik2005JPhysB,Bilous2020,Peik2021} will be sensitive to both the strong and electromagnetic interactions and undoubtedly open enormously interesting possibilities for the study of new physics, and for the establishment of ultra-stable frequency references.  

\begin{acknowledgments}
This work is part of the ``Thorium Nuclear Clock'' project that has received funding from the European Research Council under the European Union's Horizon 2020 research and innovation program (Grant No. 856415).  The calculations in this work were done through the use of Information Technologies resources at the University of Delaware, specifically the high-performance Caviness and DARWIN computer clusters.  This work was supported by National Science Foundation Grants No.~PHY-2110102,  No.~PHY-2309254, and Office of Naval Research Grants No.~N00014-22-1-2070 and N00014-20-1-2513. The authors acknowledge the support of the 20FUN01 TSCAC project. The project 23FUN03 HIOC has received funding from the European Partnership on Metrology, co-financed from the European Union's Horizon Europe Research and Innovation Programme and by the Participating States. This work has been supported by the Max Planck Society; the Max--Planck--Riken--PTB--Center for Time, Constants, and Fundamental Symmetries; and the German Federal Ministry of Education and Research (BMBF) through program grant No. 13N15973 (Projekt VAUQSI).
\end{acknowledgments}


\begin{thebibliography}{59}%
\makeatletter
\providecommand \@ifxundefined [1]{%
 \@ifx{#1\undefined}
}%
\providecommand \@ifnum [1]{%
 \ifnum #1\expandafter \@firstoftwo
 \else \expandafter \@secondoftwo
 \fi
}%
\providecommand \@ifx [1]{%
 \ifx #1\expandafter \@firstoftwo
 \else \expandafter \@secondoftwo
 \fi
}%
\providecommand \natexlab [1]{#1}%
\providecommand \enquote  [1]{``#1''}%
\providecommand \bibnamefont  [1]{#1}%
\providecommand \bibfnamefont [1]{#1}%
\providecommand \citenamefont [1]{#1}%
\providecommand \href@noop [0]{\@secondoftwo}%
\providecommand \href [0]{\begingroup \@sanitize@url \@href}%
\providecommand \@href[1]{\@@startlink{#1}\@@href}%
\providecommand \@@href[1]{\endgroup#1\@@endlink}%
\providecommand \@sanitize@url [0]{\catcode `\\12\catcode `\$12\catcode
  `\&12\catcode `\#12\catcode `\^12\catcode `\_12\catcode `\%12\relax}%
\providecommand \@@startlink[1]{}%
\providecommand \@@endlink[0]{}%
\providecommand \url  [0]{\begingroup\@sanitize@url \@url }%
\providecommand \@url [1]{\endgroup\@href {#1}{\urlprefix }}%
\providecommand \urlprefix  [0]{URL }%
\providecommand \Eprint [0]{\href }%
\providecommand \doibase [0]{http://dx.doi.org/}%
\providecommand \selectlanguage [0]{\@gobble}%
\providecommand \bibinfo  [0]{\@secondoftwo}%
\providecommand \bibfield  [0]{\@secondoftwo}%
\providecommand \translation [1]{[#1]}%
\providecommand \BibitemOpen [0]{}%
\providecommand \bibitemStop [0]{}%
\providecommand \bibitemNoStop [0]{.\EOS\space}%
\providecommand \EOS [0]{\spacefactor3000\relax}%
\providecommand \BibitemShut  [1]{\csname bibitem#1\endcsname}%
\let\auto@bib@innerbib\@empty
\bibitem [{\citenamefont {Safronova}\ \emph {et~al.}(2018)\citenamefont
  {Safronova}, \citenamefont {Budker}, \citenamefont {DeMille}, \citenamefont
  {Kimball}, \citenamefont {Derevianko},\ and\ \citenamefont
  {Clark}}]{Safronova2018RMP}%
  \BibitemOpen
  \bibfield  {author} {\bibinfo {author} {\bibfnamefont {M.~S.}\ \bibnamefont
  {Safronova}}, \bibinfo {author} {\bibfnamefont {D.}~\bibnamefont {Budker}},
  \bibinfo {author} {\bibfnamefont {D.}~\bibnamefont {DeMille}}, \bibinfo
  {author} {\bibfnamefont {D.~F.~J.}\ \bibnamefont {Kimball}}, \bibinfo
  {author} {\bibfnamefont {A.}~\bibnamefont {Derevianko}}, \ and\ \bibinfo
  {author} {\bibfnamefont {C.~W.}\ \bibnamefont {Clark}},\ }\href {\doibase
  10.1103/RevModPhys.90.025008} {\bibfield  {journal} {\bibinfo  {journal}
  {Rev. Mod. Phys.}\ }\textbf {\bibinfo {volume} {90}},\ \bibinfo {pages}
  {025008} (\bibinfo {year} {2018})}\BibitemShut {NoStop}%
\bibitem [{\citenamefont {Chou}\ \emph {et~al.}(2010)\citenamefont {Chou},
  \citenamefont {Hume}, \citenamefont {Rosenband},\ and\ \citenamefont
  {Wineland}}]{Chou2010Science}%
  \BibitemOpen
  \bibfield  {author} {\bibinfo {author} {\bibfnamefont {C.~W.}\ \bibnamefont
  {Chou}}, \bibinfo {author} {\bibfnamefont {D.~B.}\ \bibnamefont {Hume}},
  \bibinfo {author} {\bibfnamefont {T.}~\bibnamefont {Rosenband}}, \ and\
  \bibinfo {author} {\bibfnamefont {D.~J.}\ \bibnamefont {Wineland}},\
  }\href@noop {} {\bibfield  {journal} {\bibinfo  {journal} {Science}\ }\textbf
  {\bibinfo {volume} {329}},\ \bibinfo {pages} {1630} (\bibinfo {year}
  {2010})}\BibitemShut {NoStop}%
\bibitem [{\citenamefont {Bothwell}\ \emph {et~al.}(2022)\citenamefont
  {Bothwell}, \citenamefont {Kennedy}, \citenamefont {Aeppli}, \citenamefont
  {Kedar}, \citenamefont {Robinson}, \citenamefont {Oelker}, \citenamefont
  {Staron},\ and\ \citenamefont {Ye}}]{Bothwell2022Nature}%
  \BibitemOpen
  \bibfield  {author} {\bibinfo {author} {\bibfnamefont {T.}~\bibnamefont
  {Bothwell}}, \bibinfo {author} {\bibfnamefont {C.~J.}\ \bibnamefont
  {Kennedy}}, \bibinfo {author} {\bibfnamefont {A.}~\bibnamefont {Aeppli}},
  \bibinfo {author} {\bibfnamefont {D.}~\bibnamefont {Kedar}}, \bibinfo
  {author} {\bibfnamefont {J.~M.}\ \bibnamefont {Robinson}}, \bibinfo {author}
  {\bibfnamefont {E.}~\bibnamefont {Oelker}}, \bibinfo {author} {\bibfnamefont
  {A.}~\bibnamefont {Staron}}, \ and\ \bibinfo {author} {\bibfnamefont
  {J.}~\bibnamefont {Ye}},\ }\href@noop {} {\bibfield  {journal} {\bibinfo
  {journal} {Nature}\ }\textbf {\bibinfo {volume} {602}},\ \bibinfo {pages}
  {420} (\bibinfo {year} {2022})}\BibitemShut {NoStop}%
\bibitem [{\citenamefont {Zheng}\ \emph {et~al.}(2022)\citenamefont {Zheng},
  \citenamefont {Dolde}, \citenamefont {Lochab}, \citenamefont {Merriman},
  \citenamefont {Li},\ and\ \citenamefont {Kolkowitz}}]{Zheng2022Nature}%
  \BibitemOpen
  \bibfield  {author} {\bibinfo {author} {\bibfnamefont {X.}~\bibnamefont
  {Zheng}}, \bibinfo {author} {\bibfnamefont {J.}~\bibnamefont {Dolde}},
  \bibinfo {author} {\bibfnamefont {V.}~\bibnamefont {Lochab}}, \bibinfo
  {author} {\bibfnamefont {B.~N.}\ \bibnamefont {Merriman}}, \bibinfo {author}
  {\bibfnamefont {H.}~\bibnamefont {Li}}, \ and\ \bibinfo {author}
  {\bibfnamefont {S.}~\bibnamefont {Kolkowitz}},\ }\href@noop {} {\bibfield
  {journal} {\bibinfo  {journal} {Nature}\ }\textbf {\bibinfo {volume} {602}},\
  \bibinfo {pages} {425} (\bibinfo {year} {2022})}\BibitemShut {NoStop}%
\bibitem [{\citenamefont {Beloy}\ \emph {et~al.}(2021)\citenamefont {Beloy},
  \citenamefont {Bodine}, \citenamefont {Bothwell}, \citenamefont {Brewer},
  \citenamefont {Bromley}, \citenamefont {Chen}, \citenamefont {Desch\^enes},
  \citenamefont {Diddams}, \citenamefont {Fasano}, \citenamefont {Fortier},
  \citenamefont {Hassan}, \citenamefont {Hume}, \citenamefont {Kedar},
  \citenamefont {Kennedy}, \citenamefont {Khader}, \citenamefont {Koepke},
  \citenamefont {Leibrandt}, \citenamefont {Leopardi}, \citenamefont {Ludlow},
  \citenamefont {McGrew}, \citenamefont {Milner}, \citenamefont {Newbury},
  \citenamefont {Nicolodi}, \citenamefont {Oelker}, \citenamefont {Parker},
  \citenamefont {Robinson}, \citenamefont {Romisch}, \citenamefont
  {Sch\"affer}, \citenamefont {Sherman}, \citenamefont {Sonderhouse},
  \citenamefont {Swann}, \citenamefont {Yao}, \citenamefont {Ye},\ and\
  \citenamefont {Zhang}}]{BACON2021Nature}%
  \BibitemOpen
  \bibfield  {author} {\bibinfo {author} {\bibfnamefont {K.}~\bibnamefont
  {Beloy}}, \bibinfo {author} {\bibfnamefont {M.~I.}\ \bibnamefont {Bodine}},
  \bibinfo {author} {\bibfnamefont {T.}~\bibnamefont {Bothwell}}, \bibinfo
  {author} {\bibfnamefont {S.~M.}\ \bibnamefont {Brewer}}, \bibinfo {author}
  {\bibfnamefont {S.~L.}\ \bibnamefont {Bromley}}, \bibinfo {author}
  {\bibfnamefont {J.-S.}\ \bibnamefont {Chen}}, \bibinfo {author}
  {\bibfnamefont {J.-D.}\ \bibnamefont {Desch\^enes}}, \bibinfo {author}
  {\bibfnamefont {S.~A.}\ \bibnamefont {Diddams}}, \bibinfo {author}
  {\bibfnamefont {R.~J.}\ \bibnamefont {Fasano}}, \bibinfo {author}
  {\bibfnamefont {T.~M.}\ \bibnamefont {Fortier}}, \bibinfo {author}
  {\bibfnamefont {Y.~S.}\ \bibnamefont {Hassan}}, \bibinfo {author}
  {\bibfnamefont {D.~B.}\ \bibnamefont {Hume}}, \bibinfo {author}
  {\bibfnamefont {D.}~\bibnamefont {Kedar}}, \bibinfo {author} {\bibfnamefont
  {C.~J.}\ \bibnamefont {Kennedy}}, \bibinfo {author} {\bibfnamefont
  {I.}~\bibnamefont {Khader}}, \bibinfo {author} {\bibfnamefont
  {A.}~\bibnamefont {Koepke}}, \bibinfo {author} {\bibfnamefont {D.~R.}\
  \bibnamefont {Leibrandt}}, \bibinfo {author} {\bibfnamefont {H.}~\bibnamefont
  {Leopardi}}, \bibinfo {author} {\bibfnamefont {A.~D.}\ \bibnamefont
  {Ludlow}}, \bibinfo {author} {\bibfnamefont {W.~F.}\ \bibnamefont {McGrew}},
  \bibinfo {author} {\bibfnamefont {W.~R.}\ \bibnamefont {Milner}}, \bibinfo
  {author} {\bibfnamefont {N.~R.}\ \bibnamefont {Newbury}}, \bibinfo {author}
  {\bibfnamefont {D.}~\bibnamefont {Nicolodi}}, \bibinfo {author}
  {\bibfnamefont {E.}~\bibnamefont {Oelker}}, \bibinfo {author} {\bibfnamefont
  {T.~E.}\ \bibnamefont {Parker}}, \bibinfo {author} {\bibfnamefont {J.~M.}\
  \bibnamefont {Robinson}}, \bibinfo {author} {\bibfnamefont {S.}~\bibnamefont
  {Romisch}}, \bibinfo {author} {\bibfnamefont {S.~A.}\ \bibnamefont
  {Sch\"affer}}, \bibinfo {author} {\bibfnamefont {L.~C.}\ \bibnamefont
  {Sherman}, \bibfnamefont {J.~A.~Sinclair}}, \bibinfo {author} {\bibfnamefont
  {L.}~\bibnamefont {Sonderhouse}}, \bibinfo {author} {\bibfnamefont {W.~C.}\
  \bibnamefont {Swann}}, \bibinfo {author} {\bibfnamefont {J.}~\bibnamefont
  {Yao}}, \bibinfo {author} {\bibfnamefont {J.}~\bibnamefont {Ye}}, \ and\
  \bibinfo {author} {\bibfnamefont {X.}~\bibnamefont {Zhang}} (\bibinfo
  {collaboration} {Boulder Atomic Clock Optical Network (BACON)
  Collaboration}),\ }\href@noop {} {\bibfield  {journal} {\bibinfo  {journal}
  {Nature}\ }\textbf {\bibinfo {volume} {591}},\ \bibinfo {pages} {564}
  (\bibinfo {year} {2021})}\BibitemShut {NoStop}%
\bibitem [{\citenamefont {Sanner}\ \emph {et~al.}(2019)\citenamefont {Sanner},
  \citenamefont {Huntemann}, \citenamefont {Lange}, \citenamefont {Tamm},
  \citenamefont {Peik}, \citenamefont {Safronova},\ and\ \citenamefont
  {Porsev}}]{Sanner2019Nature}%
  \BibitemOpen
  \bibfield  {author} {\bibinfo {author} {\bibfnamefont {C.}~\bibnamefont
  {Sanner}}, \bibinfo {author} {\bibfnamefont {N.}~\bibnamefont {Huntemann}},
  \bibinfo {author} {\bibfnamefont {R.}~\bibnamefont {Lange}}, \bibinfo
  {author} {\bibfnamefont {C.}~\bibnamefont {Tamm}}, \bibinfo {author}
  {\bibfnamefont {E.}~\bibnamefont {Peik}}, \bibinfo {author} {\bibfnamefont
  {M.~S.}\ \bibnamefont {Safronova}}, \ and\ \bibinfo {author} {\bibfnamefont
  {S.~G.}\ \bibnamefont {Porsev}},\ }\href@noop {} {\bibfield  {journal}
  {\bibinfo  {journal} {Nature}\ }\textbf {\bibinfo {volume} {567}},\ \bibinfo
  {pages} {204} (\bibinfo {year} {2019})}\BibitemShut {NoStop}%
\bibitem [{\citenamefont {Lange}\ \emph {et~al.}(2021)\citenamefont {Lange},
  \citenamefont {Huntemann}, \citenamefont {Rahm}, \citenamefont {Sanner},
  \citenamefont {Shao}, \citenamefont {Lipphardt}, \citenamefont {Tamm},
  \citenamefont {Weyers},\ and\ \citenamefont {Peik}}]{Lange2021PRL}%
  \BibitemOpen
  \bibfield  {author} {\bibinfo {author} {\bibfnamefont {R.}~\bibnamefont
  {Lange}}, \bibinfo {author} {\bibfnamefont {N.}~\bibnamefont {Huntemann}},
  \bibinfo {author} {\bibfnamefont {J.~M.}\ \bibnamefont {Rahm}}, \bibinfo
  {author} {\bibfnamefont {C.}~\bibnamefont {Sanner}}, \bibinfo {author}
  {\bibfnamefont {H.}~\bibnamefont {Shao}}, \bibinfo {author} {\bibfnamefont
  {B.}~\bibnamefont {Lipphardt}}, \bibinfo {author} {\bibfnamefont
  {C.}~\bibnamefont {Tamm}}, \bibinfo {author} {\bibfnamefont {S.}~\bibnamefont
  {Weyers}}, \ and\ \bibinfo {author} {\bibfnamefont {E.}~\bibnamefont
  {Peik}},\ }\href {\doibase 10.1103/PhysRevLett.126.011102} {\bibfield
  {journal} {\bibinfo  {journal} {Phys. Rev. Lett.}\ }\textbf {\bibinfo
  {volume} {126}},\ \bibinfo {pages} {011102} (\bibinfo {year}
  {2021})}\BibitemShut {NoStop}%
\bibitem [{\citenamefont {Rosenband}\ \emph {et~al.}(2008)\citenamefont
  {Rosenband}, \citenamefont {Hume}, \citenamefont {Schmidt}, \citenamefont
  {Chou}, \citenamefont {Brusch}, \citenamefont {Oskay}, \citenamefont
  {Drullinger}, \citenamefont {M.}, \citenamefont {Stalnaker}, \citenamefont
  {Diddams}, \citenamefont {Swann}, \citenamefont {Newbury}, \citenamefont
  {Itano}, \citenamefont {Wineland},\ and\ \citenamefont
  {Bergquist}}]{Rosenband2008Science}%
  \BibitemOpen
  \bibfield  {author} {\bibinfo {author} {\bibfnamefont {T.}~\bibnamefont
  {Rosenband}}, \bibinfo {author} {\bibfnamefont {D.~B.}\ \bibnamefont {Hume}},
  \bibinfo {author} {\bibfnamefont {P.~O.}\ \bibnamefont {Schmidt}}, \bibinfo
  {author} {\bibfnamefont {C.~W.}\ \bibnamefont {Chou}}, \bibinfo {author}
  {\bibfnamefont {A.}~\bibnamefont {Brusch}}, \bibinfo {author} {\bibfnamefont
  {W.~H.}\ \bibnamefont {Oskay}}, \bibinfo {author} {\bibfnamefont {R.~E.}\
  \bibnamefont {Drullinger}}, \bibinfo {author} {\bibfnamefont {F.~T.}\
  \bibnamefont {M.}}, \bibinfo {author} {\bibfnamefont {J.~E.}\ \bibnamefont
  {Stalnaker}}, \bibinfo {author} {\bibfnamefont {S.~A.}\ \bibnamefont
  {Diddams}}, \bibinfo {author} {\bibfnamefont {W.~C.}\ \bibnamefont {Swann}},
  \bibinfo {author} {\bibfnamefont {N.~R.}\ \bibnamefont {Newbury}}, \bibinfo
  {author} {\bibfnamefont {W.~M.}\ \bibnamefont {Itano}}, \bibinfo {author}
  {\bibfnamefont {D.~J.}\ \bibnamefont {Wineland}}, \ and\ \bibinfo {author}
  {\bibfnamefont {J.~C.}\ \bibnamefont {Bergquist}},\ }\href@noop {} {\bibfield
   {journal} {\bibinfo  {journal} {Science}\ }\textbf {\bibinfo {volume}
  {319}},\ \bibinfo {pages} {1808} (\bibinfo {year} {2008})}\BibitemShut
  {NoStop}%
\bibitem [{\citenamefont {Godun}\ \emph {et~al.}(2014)\citenamefont {Godun},
  \citenamefont {Nisbet-Jones}, \citenamefont {Jones}, \citenamefont {King},
  \citenamefont {Johnson}, \citenamefont {Margolis}, \citenamefont {Szymaniec},
  , \citenamefont {Lea}, \citenamefont {Bongs},\ and\ \citenamefont
  {Gill}}]{Godun2014PRL}%
  \BibitemOpen
  \bibfield  {author} {\bibinfo {author} {\bibfnamefont {R.~M.}\ \bibnamefont
  {Godun}}, \bibinfo {author} {\bibfnamefont {P.~B.~R.}\ \bibnamefont
  {Nisbet-Jones}}, \bibinfo {author} {\bibfnamefont {J.~M.}\ \bibnamefont
  {Jones}}, \bibinfo {author} {\bibfnamefont {S.}~\bibnamefont {King}},
  \bibinfo {author} {\bibfnamefont {L.~A.~M.}\ \bibnamefont {Johnson}},
  \bibinfo {author} {\bibfnamefont {H.~S.}\ \bibnamefont {Margolis}}, \bibinfo
  {author} {\bibfnamefont {K.}~\bibnamefont {Szymaniec}}, , \bibinfo {author}
  {\bibfnamefont {S.~N.}\ \bibnamefont {Lea}}, \bibinfo {author} {\bibfnamefont
  {K.}~\bibnamefont {Bongs}}, \ and\ \bibinfo {author} {\bibfnamefont
  {P.}~\bibnamefont {Gill}},\ }\href@noop {} {\bibfield  {journal} {\bibinfo
  {journal} {Phys. Rev. Lett.}\ }\textbf {\bibinfo {volume} {113}},\ \bibinfo
  {pages} {210801} (\bibinfo {year} {2014})}\BibitemShut {NoStop}%
\bibitem [{\citenamefont {Huntemann}\ \emph {et~al.}(2014)\citenamefont
  {Huntemann}, \citenamefont {Lipphardt}, \citenamefont {Tamm}, \citenamefont
  {Gerginov}, \citenamefont {Weyers},\ and\ \citenamefont
  {Peik}}]{Huntemann2014PRL}%
  \BibitemOpen
  \bibfield  {author} {\bibinfo {author} {\bibfnamefont {N.}~\bibnamefont
  {Huntemann}}, \bibinfo {author} {\bibfnamefont {B.}~\bibnamefont
  {Lipphardt}}, \bibinfo {author} {\bibfnamefont {C.}~\bibnamefont {Tamm}},
  \bibinfo {author} {\bibfnamefont {V.}~\bibnamefont {Gerginov}}, \bibinfo
  {author} {\bibfnamefont {S.}~\bibnamefont {Weyers}}, \ and\ \bibinfo {author}
  {\bibfnamefont {E.}~\bibnamefont {Peik}},\ }\href@noop {} {\bibfield
  {journal} {\bibinfo  {journal} {Phys. Rev. Lett.}\ }\textbf {\bibinfo
  {volume} {113}},\ \bibinfo {pages} {210802} (\bibinfo {year}
  {2014})}\BibitemShut {NoStop}%
\bibitem [{\citenamefont {Filzinger}\ \emph {et~al.}(2023)\citenamefont
  {Filzinger}, \citenamefont {D\"orscher}, \citenamefont {Lange}, \citenamefont
  {Klose}, \citenamefont {Steinel}, \citenamefont {Benkler}, \citenamefont
  {Peik}, \citenamefont {Lisdat},\ and\ \citenamefont
  {Huntemann}}]{Filzinger2023PRL}%
  \BibitemOpen
  \bibfield  {author} {\bibinfo {author} {\bibfnamefont {M.}~\bibnamefont
  {Filzinger}}, \bibinfo {author} {\bibfnamefont {S.}~\bibnamefont
  {D\"orscher}}, \bibinfo {author} {\bibfnamefont {R.}~\bibnamefont {Lange}},
  \bibinfo {author} {\bibfnamefont {J.}~\bibnamefont {Klose}}, \bibinfo
  {author} {\bibfnamefont {M.}~\bibnamefont {Steinel}}, \bibinfo {author}
  {\bibfnamefont {E.}~\bibnamefont {Benkler}}, \bibinfo {author} {\bibfnamefont
  {E.}~\bibnamefont {Peik}}, \bibinfo {author} {\bibfnamefont {C.}~\bibnamefont
  {Lisdat}}, \ and\ \bibinfo {author} {\bibfnamefont {N.}~\bibnamefont
  {Huntemann}},\ }\href {\doibase 10.1103/PhysRevLett.130.253001} {\bibfield
  {journal} {\bibinfo  {journal} {Phys. Rev. Lett.}\ }\textbf {\bibinfo
  {volume} {130}},\ \bibinfo {pages} {253001} (\bibinfo {year}
  {2023})}\BibitemShut {NoStop}%
\bibitem [{\citenamefont {Ludlow}\ \emph {et~al.}(2015)\citenamefont {Ludlow},
  \citenamefont {Boyd}, \citenamefont {Ye}, \citenamefont {Peik},\ and\
  \citenamefont {Schmidt}}]{Ludlow2015RMP}%
  \BibitemOpen
  \bibfield  {author} {\bibinfo {author} {\bibfnamefont {A.~D.}\ \bibnamefont
  {Ludlow}}, \bibinfo {author} {\bibfnamefont {M.~M.}\ \bibnamefont {Boyd}},
  \bibinfo {author} {\bibfnamefont {J.}~\bibnamefont {Ye}}, \bibinfo {author}
  {\bibfnamefont {E.}~\bibnamefont {Peik}}, \ and\ \bibinfo {author}
  {\bibfnamefont {P.~O.}\ \bibnamefont {Schmidt}},\ }\href@noop {} {\bibfield
  {journal} {\bibinfo  {journal} {Rev. Mod. Phys.}\ }\textbf {\bibinfo {volume}
  {87}},\ \bibinfo {pages} {637} (\bibinfo {year} {2015})}\BibitemShut
  {NoStop}%
\bibitem [{\citenamefont {Kozlov}\ \emph {et~al.}(2018)\citenamefont {Kozlov},
  \citenamefont {Safronova}, \citenamefont {{Crespo L\'opez-Urrutia}},\ and\
  \citenamefont {Schmidt}}]{Kozlov2018RMP}%
  \BibitemOpen
  \bibfield  {author} {\bibinfo {author} {\bibfnamefont {M.~G.}\ \bibnamefont
  {Kozlov}}, \bibinfo {author} {\bibfnamefont {M.~S.}\ \bibnamefont
  {Safronova}}, \bibinfo {author} {\bibfnamefont {J.~R.}\ \bibnamefont {{Crespo
  L\'opez-Urrutia}}}, \ and\ \bibinfo {author} {\bibfnamefont {P.~O.}\
  \bibnamefont {Schmidt}},\ }\href {\doibase 10.1103/RevModPhys.90.045005}
  {\bibfield  {journal} {\bibinfo  {journal} {Rev. Mod. Phys.}\ }\textbf
  {\bibinfo {volume} {90}},\ \bibinfo {pages} {045005} (\bibinfo {year}
  {2018})}\BibitemShut {NoStop}%
\bibitem [{\citenamefont {Zhang}\ \emph {et~al.}(2024)\citenamefont {Zhang},
  \citenamefont {Ooi}, \citenamefont {Higgins}, \citenamefont {Doyle},
  \citenamefont {von~der Wense}, \citenamefont {Beeks}, \citenamefont
  {Leitner}, \citenamefont {Kazakov}, \citenamefont {Li}, \citenamefont
  {Thirolf}, \citenamefont {Schumm},\ and\ \citenamefont {Ye}}]{Zhang24}%
  \BibitemOpen
  \bibfield  {author} {\bibinfo {author} {\bibfnamefont {C.}~\bibnamefont
  {Zhang}}, \bibinfo {author} {\bibfnamefont {T.}~\bibnamefont {Ooi}}, \bibinfo
  {author} {\bibfnamefont {J.~S.}\ \bibnamefont {Higgins}}, \bibinfo {author}
  {\bibfnamefont {J.~F.}\ \bibnamefont {Doyle}}, \bibinfo {author}
  {\bibfnamefont {L.}~\bibnamefont {von~der Wense}}, \bibinfo {author}
  {\bibfnamefont {K.}~\bibnamefont {Beeks}}, \bibinfo {author} {\bibfnamefont
  {A.}~\bibnamefont {Leitner}}, \bibinfo {author} {\bibfnamefont
  {G.}~\bibnamefont {Kazakov}}, \bibinfo {author} {\bibfnamefont
  {P.}~\bibnamefont {Li}}, \bibinfo {author} {\bibfnamefont {P.~G.}\
  \bibnamefont {Thirolf}}, \bibinfo {author} {\bibfnamefont {T.}~\bibnamefont
  {Schumm}}, \ and\ \bibinfo {author} {\bibfnamefont {J.}~\bibnamefont {Ye}},\
  }\href {https://arxiv.org/abs/2406.18719} {\enquote {\bibinfo {title} {Dawn
  of a nuclear clock: frequency ratio of the $^{229m}${Th} isomeric transition
  and the $^{87}${Sr} atomic clock},}\ } (\bibinfo {year} {2024}),\ \Eprint
  {http://arxiv.org/abs/2406.18719} {arXiv:2406.18719 [physics.atom-ph]}
  \BibitemShut {NoStop}%
\bibitem [{\citenamefont {Berengut}\ \emph {et~al.}(2010)\citenamefont
  {Berengut}, \citenamefont {Dzuba},\ and\ \citenamefont
  {Flambaum}}]{Berengut2010PRL}%
  \BibitemOpen
  \bibfield  {author} {\bibinfo {author} {\bibfnamefont {J.~C.}\ \bibnamefont
  {Berengut}}, \bibinfo {author} {\bibfnamefont {V.~A.}\ \bibnamefont {Dzuba}},
  \ and\ \bibinfo {author} {\bibfnamefont {V.~V.}\ \bibnamefont {Flambaum}},\
  }\href {\doibase 10.1103/PhysRevLett.105.120801} {\bibfield  {journal}
  {\bibinfo  {journal} {Phys. Rev. Lett.}\ }\textbf {\bibinfo {volume} {105}},\
  \bibinfo {pages} {120801} (\bibinfo {year} {2010})}\BibitemShut {NoStop}%
\bibitem [{\citenamefont {Berengut}\ \emph {et~al.}(2011)\citenamefont
  {Berengut}, \citenamefont {Dzuba}, \citenamefont {Flambaum},\ and\
  \citenamefont {Ong}}]{berengut_electron-hole_2011}%
  \BibitemOpen
  \bibfield  {author} {\bibinfo {author} {\bibfnamefont {J.~C.}\ \bibnamefont
  {Berengut}}, \bibinfo {author} {\bibfnamefont {V.~A.}\ \bibnamefont {Dzuba}},
  \bibinfo {author} {\bibfnamefont {V.~V.}\ \bibnamefont {Flambaum}}, \ and\
  \bibinfo {author} {\bibfnamefont {A.}~\bibnamefont {Ong}},\ }\href {\doibase
  10.1103/PhysRevLett.106.210802} {\bibfield  {journal} {\bibinfo  {journal}
  {Phys. Rev. Lett.}\ }\textbf {\bibinfo {volume} {106}},\ \bibinfo {pages}
  {210802} (\bibinfo {year} {2011})}\BibitemShut {NoStop}%
\bibitem [{\citenamefont {Berengut}\ \emph {et~al.}(2012)\citenamefont
  {Berengut}, \citenamefont {Dzuba}, \citenamefont {Flambaum},\ and\
  \citenamefont {Ong}}]{berengut2012a}%
  \BibitemOpen
  \bibfield  {author} {\bibinfo {author} {\bibfnamefont {J.~C.}\ \bibnamefont
  {Berengut}}, \bibinfo {author} {\bibfnamefont {V.~A.}\ \bibnamefont {Dzuba}},
  \bibinfo {author} {\bibfnamefont {V.~V.}\ \bibnamefont {Flambaum}}, \ and\
  \bibinfo {author} {\bibfnamefont {A.}~\bibnamefont {Ong}},\ }\href {\doibase
  10.1103/PhysRevA.86.022517} {\bibfield  {journal} {\bibinfo  {journal} {Phys.
  Rev. A}\ }\textbf {\bibinfo {volume} {86}},\ \bibinfo {pages} {022517}
  (\bibinfo {year} {2012})}\BibitemShut {NoStop}%
\bibitem [{\citenamefont {Safronova}\ \emph
  {et~al.}(2014{\natexlab{a}})\citenamefont {Safronova}, \citenamefont {Dzuba},
  \citenamefont {Flambaum}, \citenamefont {Safronova}, \citenamefont {Porsev},\
  and\ \citenamefont {Kozlov}}]{Safronova2014PRL}%
  \BibitemOpen
  \bibfield  {author} {\bibinfo {author} {\bibfnamefont {M.~S.}\ \bibnamefont
  {Safronova}}, \bibinfo {author} {\bibfnamefont {V.~A.}\ \bibnamefont
  {Dzuba}}, \bibinfo {author} {\bibfnamefont {V.~V.}\ \bibnamefont {Flambaum}},
  \bibinfo {author} {\bibfnamefont {U.~I.}\ \bibnamefont {Safronova}}, \bibinfo
  {author} {\bibfnamefont {S.~G.}\ \bibnamefont {Porsev}}, \ and\ \bibinfo
  {author} {\bibfnamefont {M.~G.}\ \bibnamefont {Kozlov}},\ }\href@noop {}
  {\bibfield  {journal} {\bibinfo  {journal} {Phys. Rev. Lett.}\ }\textbf
  {\bibinfo {volume} {113}},\ \bibinfo {pages} {030801} (\bibinfo {year}
  {2014}{\natexlab{a}})}\BibitemShut {NoStop}%
\bibitem [{\citenamefont {Ong}\ \emph {et~al.}(2014)\citenamefont {Ong},
  \citenamefont {Berengut},\ and\ \citenamefont {Flambaum}}]{ong2014optical}%
  \BibitemOpen
  \bibfield  {author} {\bibinfo {author} {\bibfnamefont {A.}~\bibnamefont
  {Ong}}, \bibinfo {author} {\bibfnamefont {J.~C.}\ \bibnamefont {Berengut}}, \
  and\ \bibinfo {author} {\bibfnamefont {V.~V.}\ \bibnamefont {Flambaum}},\
  }\enquote {\bibinfo {title} {Optical transitions in highly charged ions for
  detection of variations in the fine-structure constant},}\ in\ \href
  {\doibase 10.1007/978-3-642-45201-7_9} {\emph {\bibinfo {booktitle}
  {Fundamental Physics in Particle Traps}}},\ \bibinfo {editor} {edited by\
  \bibinfo {editor} {\bibfnamefont {W.}~\bibnamefont {Quint}}\ and\ \bibinfo
  {editor} {\bibfnamefont {M.}~\bibnamefont {Vogel}}}\ (\bibinfo  {publisher}
  {Springer Berlin Heidelberg},\ \bibinfo {address} {Berlin, Heidelberg},\
  \bibinfo {year} {2014})\ pp.\ \bibinfo {pages} {293--314}\BibitemShut
  {NoStop}%
\bibitem [{\citenamefont {Dzuba}\ and\ \citenamefont
  {Flambaum}(2015)}]{Dzuba2015}%
  \BibitemOpen
  \bibfield  {author} {\bibinfo {author} {\bibfnamefont {V.~A.}\ \bibnamefont
  {Dzuba}}\ and\ \bibinfo {author} {\bibfnamefont {V.~V.}\ \bibnamefont
  {Flambaum}},\ }\href {\doibase 10.1007/s10751-015-1166-4} {\bibfield
  {journal} {\bibinfo  {journal} {Hyperfine Interact.}\ }\textbf {\bibinfo
  {volume} {236}},\ \bibinfo {pages} {79} (\bibinfo {year} {2015})}\BibitemShut
  {NoStop}%
\bibitem [{\citenamefont {Schiller}(2007)}]{Schiller2007PRL}%
  \BibitemOpen
  \bibfield  {author} {\bibinfo {author} {\bibfnamefont {S.}~\bibnamefont
  {Schiller}},\ }\href@noop {} {\bibfield  {journal} {\bibinfo  {journal}
  {Phys. Rev. Lett.}\ }\textbf {\bibinfo {volume} {98}},\ \bibinfo {pages}
  {180801} (\bibinfo {year} {2007})}\BibitemShut {NoStop}%
\bibitem [{\citenamefont {Oreshkina}\ \emph {et~al.}(2017)\citenamefont
  {Oreshkina}, \citenamefont {Cavaletto}, \citenamefont {Michel}, \citenamefont
  {Harman},\ and\ \citenamefont {Keitel}}]{Oreshkina2017}%
  \BibitemOpen
  \bibfield  {author} {\bibinfo {author} {\bibfnamefont {N.~S.}\ \bibnamefont
  {Oreshkina}}, \bibinfo {author} {\bibfnamefont {S.~M.}\ \bibnamefont
  {Cavaletto}}, \bibinfo {author} {\bibfnamefont {N.}~\bibnamefont {Michel}},
  \bibinfo {author} {\bibfnamefont {Z.}~\bibnamefont {Harman}}, \ and\ \bibinfo
  {author} {\bibfnamefont {C.~H.}\ \bibnamefont {Keitel}},\ }\href {\doibase
  10.1103/PhysRevA.96.030501} {\bibfield  {journal} {\bibinfo  {journal} {Phys.
  Rev. A}\ }\textbf {\bibinfo {volume} {96}},\ \bibinfo {pages} {030501}
  (\bibinfo {year} {2017})}\BibitemShut {NoStop}%
\bibitem [{\citenamefont {Schmidt}\ \emph {et~al.}(2005)\citenamefont
  {Schmidt}, \citenamefont {Rosenband}, \citenamefont {Langer}, \citenamefont
  {Intano}, \citenamefont {Bergquist},\ and\ \citenamefont
  {Wineland}}]{Schmidt2005Science}%
  \BibitemOpen
  \bibfield  {author} {\bibinfo {author} {\bibfnamefont {P.~O.}\ \bibnamefont
  {Schmidt}}, \bibinfo {author} {\bibfnamefont {T.}~\bibnamefont {Rosenband}},
  \bibinfo {author} {\bibfnamefont {C.}~\bibnamefont {Langer}}, \bibinfo
  {author} {\bibfnamefont {W.~M.}\ \bibnamefont {Intano}}, \bibinfo {author}
  {\bibfnamefont {J.~C.}\ \bibnamefont {Bergquist}}, \ and\ \bibinfo {author}
  {\bibfnamefont {D.~J.}\ \bibnamefont {Wineland}},\ }\href@noop {} {\bibfield
  {journal} {\bibinfo  {journal} {Science}\ }\textbf {\bibinfo {volume}
  {309}},\ \bibinfo {pages} {749} (\bibinfo {year} {2005})}\BibitemShut
  {NoStop}%
\bibitem [{\citenamefont {Huntemann}\ \emph {et~al.}(2016)\citenamefont
  {Huntemann}, \citenamefont {Sanner}, \citenamefont {Lipphardt}, \citenamefont
  {Tamm},\ and\ \citenamefont {Peik}}]{Huntemann2016PRL}%
  \BibitemOpen
  \bibfield  {author} {\bibinfo {author} {\bibfnamefont {N.}~\bibnamefont
  {Huntemann}}, \bibinfo {author} {\bibfnamefont {C.}~\bibnamefont {Sanner}},
  \bibinfo {author} {\bibfnamefont {B.}~\bibnamefont {Lipphardt}}, \bibinfo
  {author} {\bibfnamefont {C.}~\bibnamefont {Tamm}}, \ and\ \bibinfo {author}
  {\bibfnamefont {E.}~\bibnamefont {Peik}},\ }\href@noop {} {\bibfield
  {journal} {\bibinfo  {journal} {Phys. Rev. Lett.}\ }\textbf {\bibinfo
  {volume} {116}},\ \bibinfo {pages} {063001} (\bibinfo {year}
  {2016})}\BibitemShut {NoStop}%
\bibitem [{\citenamefont {Bekker}\ \emph {et~al.}(2019)\citenamefont {Bekker},
  \citenamefont {Borschevsky}, \citenamefont {Harman}, \citenamefont {Keitel},
  \citenamefont {Pfeifer}, \citenamefont {Schmidt}, \citenamefont {{Crespo
  L{\'{o}}pez-Urrutia}},\ and\ \citenamefont {Berengut}}]{Bekker2019}%
  \BibitemOpen
  \bibfield  {author} {\bibinfo {author} {\bibfnamefont {H.}~\bibnamefont
  {Bekker}}, \bibinfo {author} {\bibfnamefont {A.}~\bibnamefont {Borschevsky}},
  \bibinfo {author} {\bibfnamefont {Z.}~\bibnamefont {Harman}}, \bibinfo
  {author} {\bibfnamefont {C.~H.}\ \bibnamefont {Keitel}}, \bibinfo {author}
  {\bibfnamefont {T.}~\bibnamefont {Pfeifer}}, \bibinfo {author} {\bibfnamefont
  {P.~O.}\ \bibnamefont {Schmidt}}, \bibinfo {author} {\bibfnamefont {J.~R.}\
  \bibnamefont {{Crespo L{\'{o}}pez-Urrutia}}}, \ and\ \bibinfo {author}
  {\bibfnamefont {J.~C.}\ \bibnamefont {Berengut}},\ }\href@noop {} {\bibfield
  {journal} {\bibinfo  {journal} {Nature Comm.}\ }\textbf {\bibinfo {volume}
  {10}},\ \bibinfo {pages} {5651} (\bibinfo {year} {2019})}\BibitemShut
  {NoStop}%
\bibitem [{\citenamefont {Porsev}\ \emph {et~al.}(2020)\citenamefont {Porsev},
  \citenamefont {Safronova}, \citenamefont {Safronova}, \citenamefont
  {Schmidt}, \citenamefont {Bondarev}, \citenamefont {Kozlov}, \citenamefont
  {Tupitsyn},\ and\ \citenamefont {Cheung}}]{Cf}%
  \BibitemOpen
  \bibfield  {author} {\bibinfo {author} {\bibfnamefont {S.~G.}\ \bibnamefont
  {Porsev}}, \bibinfo {author} {\bibfnamefont {U.~I.}\ \bibnamefont
  {Safronova}}, \bibinfo {author} {\bibfnamefont {M.~S.}\ \bibnamefont
  {Safronova}}, \bibinfo {author} {\bibfnamefont {P.~O.}\ \bibnamefont
  {Schmidt}}, \bibinfo {author} {\bibfnamefont {A.~I.}\ \bibnamefont
  {Bondarev}}, \bibinfo {author} {\bibfnamefont {M.~G.}\ \bibnamefont
  {Kozlov}}, \bibinfo {author} {\bibfnamefont {I.~I.}\ \bibnamefont
  {Tupitsyn}}, \ and\ \bibinfo {author} {\bibfnamefont {C.}~\bibnamefont
  {Cheung}},\ }\href {\doibase 10.1103/PhysRevA.102.012802} {\bibfield
  {journal} {\bibinfo  {journal} {Phys. Rev. A}\ }\textbf {\bibinfo {volume}
  {102}},\ \bibinfo {pages} {012802} (\bibinfo {year} {2020})}\BibitemShut
  {NoStop}%
\bibitem [{\citenamefont {Safronova}\ \emph {et~al.}(2009)\citenamefont
  {Safronova}, \citenamefont {Kozlov}, \citenamefont {Johnson},\ and\
  \citenamefont {Jiang}}]{Safronova2009}%
  \BibitemOpen
  \bibfield  {author} {\bibinfo {author} {\bibfnamefont {M.~S.}\ \bibnamefont
  {Safronova}}, \bibinfo {author} {\bibfnamefont {M.~G.}\ \bibnamefont
  {Kozlov}}, \bibinfo {author} {\bibfnamefont {W.~R.}\ \bibnamefont {Johnson}},
  \ and\ \bibinfo {author} {\bibfnamefont {D.}~\bibnamefont {Jiang}},\
  }\href@noop {} {\bibfield  {journal} {\bibinfo  {journal} {Phys. Rev. A}\
  }\textbf {\bibinfo {volume} {80}},\ \bibinfo {pages} {012516} (\bibinfo
  {year} {2009})}\BibitemShut {NoStop}%
\bibitem [{\citenamefont {Cheung}\ \emph {et~al.}(2021)\citenamefont {Cheung},
  \citenamefont {Safronova},\ and\ \citenamefont {Porsev}}]{sym2021}%
  \BibitemOpen
  \bibfield  {author} {\bibinfo {author} {\bibfnamefont {C.}~\bibnamefont
  {Cheung}}, \bibinfo {author} {\bibfnamefont {M.}~\bibnamefont {Safronova}}, \
  and\ \bibinfo {author} {\bibfnamefont {S.}~\bibnamefont {Porsev}},\ }\href
  {\doibase 10.3390/sym13040621} {\bibfield  {journal} {\bibinfo  {journal}
  {Symmetry}\ }\textbf {\bibinfo {volume} {13}},\ \bibinfo {pages} {621}
  (\bibinfo {year} {2021})}\BibitemShut {NoStop}%
\bibitem [{\citenamefont {Porsev}\ and\ \citenamefont
  {Derevianko}(2006)}]{PorDer06}%
  \BibitemOpen
  \bibfield  {author} {\bibinfo {author} {\bibfnamefont {S.~G.}\ \bibnamefont
  {Porsev}}\ and\ \bibinfo {author} {\bibfnamefont {A.}~\bibnamefont
  {Derevianko}},\ }\href@noop {} {\bibfield  {journal} {\bibinfo  {journal}
  {Phys. Rev. A}\ }\textbf {\bibinfo {volume} {73}},\ \bibinfo {pages} {012501}
  (\bibinfo {year} {2006})}\BibitemShut {NoStop}%
\bibitem [{\citenamefont {Tupitsyn}\ \emph {et~al.}(2016)\citenamefont
  {Tupitsyn}, \citenamefont {Kozlov}, \citenamefont {Safronova}, \citenamefont
  {Shabaev},\ and\ \citenamefont {Dzuba}}]{QED}%
  \BibitemOpen
  \bibfield  {author} {\bibinfo {author} {\bibfnamefont {I.~I.}\ \bibnamefont
  {Tupitsyn}}, \bibinfo {author} {\bibfnamefont {M.~G.}\ \bibnamefont
  {Kozlov}}, \bibinfo {author} {\bibfnamefont {M.~S.}\ \bibnamefont
  {Safronova}}, \bibinfo {author} {\bibfnamefont {V.~M.}\ \bibnamefont
  {Shabaev}}, \ and\ \bibinfo {author} {\bibfnamefont {V.~A.}\ \bibnamefont
  {Dzuba}},\ }\href@noop {} {\bibfield  {journal} {\bibinfo  {journal} {Phys.\
  Rev.\ Lett.}\ }\textbf {\bibinfo {volume} {117}},\ \bibinfo {pages} {253001}
  (\bibinfo {year} {2016})}\BibitemShut {NoStop}%
\bibitem [{\citenamefont {Kozlov}\ \emph {et~al.}(2016)\citenamefont {Kozlov},
  \citenamefont {Safronova}, \citenamefont {Porsev},\ and\ \citenamefont
  {Tupitsyn}}]{TEI}%
  \BibitemOpen
  \bibfield  {author} {\bibinfo {author} {\bibfnamefont {M.~G.}\ \bibnamefont
  {Kozlov}}, \bibinfo {author} {\bibfnamefont {M.~S.}\ \bibnamefont
  {Safronova}}, \bibinfo {author} {\bibfnamefont {S.~G.}\ \bibnamefont
  {Porsev}}, \ and\ \bibinfo {author} {\bibfnamefont {I.~I.}\ \bibnamefont
  {Tupitsyn}},\ }\href@noop {} {\bibfield  {journal} {\bibinfo  {journal}
  {Phys. Rev. A}\ }\textbf {\bibinfo {volume} {94}},\ \bibinfo {pages} {032512}
  (\bibinfo {year} {2016})}\BibitemShut {NoStop}%
\bibitem [{\citenamefont {Safronova}\ \emph
  {et~al.}(2014{\natexlab{b}})\citenamefont {Safronova}, \citenamefont {Dzuba},
  \citenamefont {Flambaum}, \citenamefont {Safronova}, \citenamefont {Porsev},\
  and\ \citenamefont {Kozlov}}]{SafDzuFla14b}%
  \BibitemOpen
  \bibfield  {author} {\bibinfo {author} {\bibfnamefont {M.~S.}\ \bibnamefont
  {Safronova}}, \bibinfo {author} {\bibfnamefont {V.~A.}\ \bibnamefont
  {Dzuba}}, \bibinfo {author} {\bibfnamefont {V.~V.}\ \bibnamefont {Flambaum}},
  \bibinfo {author} {\bibfnamefont {U.~I.}\ \bibnamefont {Safronova}}, \bibinfo
  {author} {\bibfnamefont {S.~G.}\ \bibnamefont {Porsev}}, \ and\ \bibinfo
  {author} {\bibfnamefont {M.~G.}\ \bibnamefont {Kozlov}},\ }\href@noop {}
  {\bibfield  {journal} {\bibinfo  {journal} {Phys. Rev. A}\ }\textbf {\bibinfo
  {volume} {90}},\ \bibinfo {pages} {042513} (\bibinfo {year}
  {2014}{\natexlab{b}})}\BibitemShut {NoStop}%
\bibitem [{\citenamefont {Stone}(2005)}]{Sto05}%
  \BibitemOpen
  \bibfield  {author} {\bibinfo {author} {\bibfnamefont {N.~J.}\ \bibnamefont
  {Stone}},\ }\href@noop {} {\bibfield  {journal} {\bibinfo  {journal} {At.
  Data Nucl. Data Tables}\ }\textbf {\bibinfo {volume} {90}},\ \bibinfo {pages}
  {75} (\bibinfo {year} {2005})}\BibitemShut {NoStop}%
\bibitem [{\citenamefont {Dzuba}\ \emph {et~al.}(1998)\citenamefont {Dzuba},
  \citenamefont {Kozlov}, \citenamefont {Porsev},\ and\ \citenamefont
  {Flambaum}}]{DzuKozPor98}%
  \BibitemOpen
  \bibfield  {author} {\bibinfo {author} {\bibfnamefont {V.~A.}\ \bibnamefont
  {Dzuba}}, \bibinfo {author} {\bibfnamefont {M.~G.}\ \bibnamefont {Kozlov}},
  \bibinfo {author} {\bibfnamefont {S.~G.}\ \bibnamefont {Porsev}}, \ and\
  \bibinfo {author} {\bibfnamefont {V.~V.}\ \bibnamefont {Flambaum}},\
  }\href@noop {} {\bibfield  {journal} {\bibinfo  {journal} {Zh. \ Eksp. \
  Teor. \ Fiz.}\ }\textbf {\bibinfo {volume} {114}},\ \bibinfo {pages} {1636}
  (\bibinfo {year} {1998})},\ \bibinfo {note} {[Sov. \ Phys.--JETP {\bf 87}
  885, (1998)]}\BibitemShut {NoStop}%
\bibitem [{\citenamefont {{Crespo L\'{o}pez-Urrutia}}\ \emph
  {et~al.}(2004)\citenamefont {{Crespo L\'{o}pez-Urrutia}}, \citenamefont
  {Braun}, \citenamefont {Brenner}, \citenamefont {Bruhns}, \citenamefont
  {Dimopoulou}, \citenamefont {Dragani\'{c}}, \citenamefont {Fischer},
  \citenamefont {Gonz\'{a}lez~Mart\'{i}nez}, \citenamefont {Lapierre},
  \citenamefont {Mironov}, \citenamefont {Moshammer}, \citenamefont {Orts},
  \citenamefont {Tawara}, \citenamefont {Trinczek},\ and\ \citenamefont
  {Ullrich}}]{crespolopez-urrutiaProgressHeidelbergEBIT2004}%
  \BibitemOpen
  \bibfield  {author} {\bibinfo {author} {\bibfnamefont {J.~R.}\ \bibnamefont
  {{Crespo L\'{o}pez-Urrutia}}}, \bibinfo {author} {\bibfnamefont
  {J.}~\bibnamefont {Braun}}, \bibinfo {author} {\bibfnamefont
  {G.}~\bibnamefont {Brenner}}, \bibinfo {author} {\bibfnamefont
  {H.}~\bibnamefont {Bruhns}}, \bibinfo {author} {\bibfnamefont
  {C.}~\bibnamefont {Dimopoulou}}, \bibinfo {author} {\bibfnamefont {I.~N.}\
  \bibnamefont {Dragani\'{c}}}, \bibinfo {author} {\bibfnamefont
  {D.}~\bibnamefont {Fischer}}, \bibinfo {author} {\bibfnamefont {A.~J.}\
  \bibnamefont {Gonz\'{a}lez~Mart\'{i}nez}}, \bibinfo {author} {\bibfnamefont
  {A.}~\bibnamefont {Lapierre}}, \bibinfo {author} {\bibfnamefont
  {V.}~\bibnamefont {Mironov}}, \bibinfo {author} {\bibfnamefont
  {R.}~\bibnamefont {Moshammer}}, \bibinfo {author} {\bibfnamefont {R.~S.}\
  \bibnamefont {Orts}}, \bibinfo {author} {\bibfnamefont {H.}~\bibnamefont
  {Tawara}}, \bibinfo {author} {\bibfnamefont {M.}~\bibnamefont {Trinczek}}, \
  and\ \bibinfo {author} {\bibfnamefont {J.}~\bibnamefont {Ullrich}},\ }\href
  {\doibase 10.1088/1742-6596/2/1/006} {\bibfield  {journal} {\bibinfo
  {journal} {J. Phys.: Conference Series}\ }\textbf {\bibinfo {volume} {2}},\
  \bibinfo {pages} {42} (\bibinfo {year} {2004})}\BibitemShut {NoStop}%
\bibitem [{\citenamefont {Levine}\ \emph {et~al.}(1988)\citenamefont {Levine},
  \citenamefont {Marrs}, \citenamefont {Henderson}, \citenamefont {Knapp},\
  and\ \citenamefont {Schneider}}]{Levine1988}%
  \BibitemOpen
  \bibfield  {author} {\bibinfo {author} {\bibfnamefont {M.~A.}\ \bibnamefont
  {Levine}}, \bibinfo {author} {\bibfnamefont {R.~E.}\ \bibnamefont {Marrs}},
  \bibinfo {author} {\bibfnamefont {J.~R.}\ \bibnamefont {Henderson}}, \bibinfo
  {author} {\bibfnamefont {D.~A.}\ \bibnamefont {Knapp}}, \ and\ \bibinfo
  {author} {\bibfnamefont {M.~B.}\ \bibnamefont {Schneider}},\ }\href {\doibase
  10.1088/0031-8949/1988/T22/024} {\bibfield  {journal} {\bibinfo  {journal}
  {Physica Scripta}\ }\textbf {\bibinfo {volume} {1988}},\ \bibinfo {pages}
  {157} (\bibinfo {year} {1988})}\BibitemShut {NoStop}%
\bibitem [{\citenamefont {Penetrante}\ \emph {et~al.}(1991)\citenamefont
  {Penetrante}, \citenamefont {Bardsley}, \citenamefont {Levine}, \citenamefont
  {Knapp},\ and\ \citenamefont {Marrs}}]{Penetrante1991}%
  \BibitemOpen
  \bibfield  {author} {\bibinfo {author} {\bibfnamefont {B.~M.}\ \bibnamefont
  {Penetrante}}, \bibinfo {author} {\bibfnamefont {J.~N.}\ \bibnamefont
  {Bardsley}}, \bibinfo {author} {\bibfnamefont {M.~A.}\ \bibnamefont
  {Levine}}, \bibinfo {author} {\bibfnamefont {D.~A.}\ \bibnamefont {Knapp}}, \
  and\ \bibinfo {author} {\bibfnamefont {R.~E.}\ \bibnamefont {Marrs}},\ }\href
  {\doibase 10.1103/PhysRevA.43.4873} {\bibfield  {journal} {\bibinfo
  {journal} {Phys. Rev. A}\ }\textbf {\bibinfo {volume} {43}},\ \bibinfo
  {pages} {4873} (\bibinfo {year} {1991})}\BibitemShut {NoStop}%
\bibitem [{\citenamefont {Bekker}\ \emph {et~al.}(2018)\citenamefont {Bekker},
  \citenamefont {Hensel}, \citenamefont {Daniel}, \citenamefont {Windberger},
  \citenamefont {Pfeifer},\ and\ \citenamefont {{Crespo
  L\'{o}pez-Urrutia}}}]{bekkerLaboratoryPrecisionMeasurements2018}%
  \BibitemOpen
  \bibfield  {author} {\bibinfo {author} {\bibfnamefont {H.}~\bibnamefont
  {Bekker}}, \bibinfo {author} {\bibfnamefont {C.}~\bibnamefont {Hensel}},
  \bibinfo {author} {\bibfnamefont {A.}~\bibnamefont {Daniel}}, \bibinfo
  {author} {\bibfnamefont {A.}~\bibnamefont {Windberger}}, \bibinfo {author}
  {\bibfnamefont {T.}~\bibnamefont {Pfeifer}}, \ and\ \bibinfo {author}
  {\bibfnamefont {J.~R.}\ \bibnamefont {{Crespo L\'{o}pez-Urrutia}}},\ }\href
  {\doibase 10.1103/PhysRevA.98.062514} {\bibfield  {journal} {\bibinfo
  {journal} {Phys. Rev. A}\ }\textbf {\bibinfo {volume} {98}},\ \bibinfo
  {pages} {062514} (\bibinfo {year} {2018})}\BibitemShut {NoStop}%
\bibitem [{\citenamefont {Tremblay}\ \emph {et~al.}(1990)\citenamefont
  {Tremblay}, \citenamefont {Michaud}, \citenamefont {Levesque}, \citenamefont
  {Th\'{e}riault}, \citenamefont {Breton}, \citenamefont {Beaubien},\ and\
  \citenamefont {Cyr}}]{tremblayAbsorptionProfilesAlkalimetal1990}%
  \BibitemOpen
  \bibfield  {author} {\bibinfo {author} {\bibfnamefont {P.}~\bibnamefont
  {Tremblay}}, \bibinfo {author} {\bibfnamefont {A.}~\bibnamefont {Michaud}},
  \bibinfo {author} {\bibfnamefont {M.}~\bibnamefont {Levesque}}, \bibinfo
  {author} {\bibfnamefont {S.}~\bibnamefont {Th\'{e}riault}}, \bibinfo {author}
  {\bibfnamefont {M.}~\bibnamefont {Breton}}, \bibinfo {author} {\bibfnamefont
  {J.}~\bibnamefont {Beaubien}}, \ and\ \bibinfo {author} {\bibfnamefont
  {N.}~\bibnamefont {Cyr}},\ }\href {\doibase 10.1103/PhysRevA.42.2766}
  {\bibfield  {journal} {\bibinfo  {journal} {Phys. Rev. A}\ }\textbf {\bibinfo
  {volume} {42}},\ \bibinfo {pages} {2766} (\bibinfo {year}
  {1990})}\BibitemShut {NoStop}%
\bibitem [{\citenamefont {Brewer}\ \emph {et~al.}(2019)\citenamefont {Brewer},
  \citenamefont {Chen}, \citenamefont {Hankin}, \citenamefont {Clements},
  \citenamefont {Chou}, \citenamefont {Wineland}, \citenamefont {Hume},\ and\
  \citenamefont {Leibrandt}}]{Brewer2019PRL}%
  \BibitemOpen
  \bibfield  {author} {\bibinfo {author} {\bibfnamefont {S.~M.}\ \bibnamefont
  {Brewer}}, \bibinfo {author} {\bibfnamefont {J.-S.}\ \bibnamefont {Chen}},
  \bibinfo {author} {\bibfnamefont {A.~M.}\ \bibnamefont {Hankin}}, \bibinfo
  {author} {\bibfnamefont {E.~R.}\ \bibnamefont {Clements}}, \bibinfo {author}
  {\bibfnamefont {C.~W.}\ \bibnamefont {Chou}}, \bibinfo {author}
  {\bibfnamefont {D.~J.}\ \bibnamefont {Wineland}}, \bibinfo {author}
  {\bibfnamefont {D.~B.}\ \bibnamefont {Hume}}, \ and\ \bibinfo {author}
  {\bibfnamefont {D.~R.}\ \bibnamefont {Leibrandt}},\ }\href {\doibase
  10.1103/PhysRevLett.123.033201} {\bibfield  {journal} {\bibinfo  {journal}
  {Phys. Rev. Lett.}\ }\textbf {\bibinfo {volume} {123}},\ \bibinfo {pages}
  {033201} (\bibinfo {year} {2019})}\BibitemShut {NoStop}%
\bibitem [{\citenamefont {Banducci}\ \emph {et~al.}()\citenamefont {Banducci},
  \citenamefont {Naing}, \citenamefont {VanArsdale},\ and\ \citenamefont
  {Brewer}}]{BanducciInPrep}%
  \BibitemOpen
  \bibfield  {author} {\bibinfo {author} {\bibfnamefont {A.~L.}\ \bibnamefont
  {Banducci}}, \bibinfo {author} {\bibfnamefont {A.~S.}\ \bibnamefont {Naing}},
  \bibinfo {author} {\bibfnamefont {J.~B.}\ \bibnamefont {VanArsdale}}, \ and\
  \bibinfo {author} {\bibfnamefont {S.~M.}\ \bibnamefont {Brewer}},\
  }\href@noop {} {\enquote {\bibinfo {title} {{Compact Electron Beam Ion trap
  for Production of Highly Charged Metal Ions}},}\ }\bibinfo {note} {(in
  preparation)}\BibitemShut {NoStop}%
\bibitem [{\citenamefont {Schm{\"o}ger}\ \emph {et~al.}(2015)\citenamefont
  {Schm{\"o}ger}, \citenamefont {Versolato}, \citenamefont {Schwarz},
  \citenamefont {Kohnen}, \citenamefont {Windberger}, \citenamefont {Piest},
  \citenamefont {Feuchtenbeiner}, \citenamefont {Pedregosa-Gutierrez},
  \citenamefont {Leopold}, \citenamefont {Micke}, \citenamefont {Hansen},
  \citenamefont {Baumann}, \citenamefont {Drewsen}, \citenamefont {Ullrich},
  \citenamefont {Schmidt},\ and\ \citenamefont {{Crespo
  L{\'o}pez-Urrutia}}}]{SchmogerScience2015}%
  \BibitemOpen
  \bibfield  {author} {\bibinfo {author} {\bibfnamefont {L.}~\bibnamefont
  {Schm{\"o}ger}}, \bibinfo {author} {\bibfnamefont {O.~O.}\ \bibnamefont
  {Versolato}}, \bibinfo {author} {\bibfnamefont {M.}~\bibnamefont {Schwarz}},
  \bibinfo {author} {\bibfnamefont {M.}~\bibnamefont {Kohnen}}, \bibinfo
  {author} {\bibfnamefont {A.}~\bibnamefont {Windberger}}, \bibinfo {author}
  {\bibfnamefont {B.}~\bibnamefont {Piest}}, \bibinfo {author} {\bibfnamefont
  {S.}~\bibnamefont {Feuchtenbeiner}}, \bibinfo {author} {\bibfnamefont
  {J.}~\bibnamefont {Pedregosa-Gutierrez}}, \bibinfo {author} {\bibfnamefont
  {T.}~\bibnamefont {Leopold}}, \bibinfo {author} {\bibfnamefont
  {P.}~\bibnamefont {Micke}}, \bibinfo {author} {\bibfnamefont {A.~K.}\
  \bibnamefont {Hansen}}, \bibinfo {author} {\bibfnamefont {T.~M.}\
  \bibnamefont {Baumann}}, \bibinfo {author} {\bibfnamefont {M.}~\bibnamefont
  {Drewsen}}, \bibinfo {author} {\bibfnamefont {J.}~\bibnamefont {Ullrich}},
  \bibinfo {author} {\bibfnamefont {P.~O.}\ \bibnamefont {Schmidt}}, \ and\
  \bibinfo {author} {\bibfnamefont {J.~R.}\ \bibnamefont {{Crespo
  L{\'o}pez-Urrutia}}},\ }\href {\doibase 10.1126/science.aaa2960} {\bibfield
  {journal} {\bibinfo  {journal} {Science}\ }\textbf {\bibinfo {volume}
  {347}},\ \bibinfo {pages} {1233} (\bibinfo {year} {2015})}\BibitemShut
  {NoStop}%
\bibitem [{\citenamefont {Micke}\ \emph {et~al.}(2020)\citenamefont {Micke},
  \citenamefont {Leopold}, \citenamefont {King}, \citenamefont {Benkler},
  \citenamefont {Spie\ss{}}, \citenamefont {Schm\"{o}ger}, \citenamefont
  {Schwarz}, \citenamefont {{Crespo L\'{o}pez-Urrutia}},\ and\ \citenamefont
  {Schmidt}}]{Micke2020Nature}%
  \BibitemOpen
  \bibfield  {author} {\bibinfo {author} {\bibfnamefont {P.}~\bibnamefont
  {Micke}}, \bibinfo {author} {\bibfnamefont {T.}~\bibnamefont {Leopold}},
  \bibinfo {author} {\bibfnamefont {S.~A.}\ \bibnamefont {King}}, \bibinfo
  {author} {\bibfnamefont {E.}~\bibnamefont {Benkler}}, \bibinfo {author}
  {\bibfnamefont {L.~J.}\ \bibnamefont {Spie\ss{}}}, \bibinfo {author}
  {\bibfnamefont {L.}~\bibnamefont {Schm\"{o}ger}}, \bibinfo {author}
  {\bibfnamefont {M.}~\bibnamefont {Schwarz}}, \bibinfo {author} {\bibfnamefont
  {J.~R.}\ \bibnamefont {{Crespo L\'{o}pez-Urrutia}}}, \ and\ \bibinfo {author}
  {\bibfnamefont {P.~O.}\ \bibnamefont {Schmidt}},\ }\href@noop {} {\bibfield
  {journal} {\bibinfo  {journal} {Nature}\ }\textbf {\bibinfo {volume} {578}},\
  \bibinfo {pages} {60} (\bibinfo {year} {2020})}\BibitemShut {NoStop}%
\bibitem [{\citenamefont {King}\ \emph {et~al.}(2022)\citenamefont {King},
  \citenamefont {Spie{\ss}}, \citenamefont {Micke}, \citenamefont {Wilzewski},
  \citenamefont {Leopold}, \citenamefont {Benkler}, \citenamefont {Lange},
  \citenamefont {Huntemann}, \citenamefont {Surzhykov}, \citenamefont
  {Yerokhin}, \citenamefont {{Crespo L{\'o}pez-Urrutia}},\ and\ \citenamefont
  {Schmidt}}]{King2022Nature}%
  \BibitemOpen
  \bibfield  {author} {\bibinfo {author} {\bibfnamefont {S.~A.}\ \bibnamefont
  {King}}, \bibinfo {author} {\bibfnamefont {L.~J.}\ \bibnamefont {Spie{\ss}}},
  \bibinfo {author} {\bibfnamefont {P.}~\bibnamefont {Micke}}, \bibinfo
  {author} {\bibfnamefont {A.}~\bibnamefont {Wilzewski}}, \bibinfo {author}
  {\bibfnamefont {T.}~\bibnamefont {Leopold}}, \bibinfo {author} {\bibfnamefont
  {E.}~\bibnamefont {Benkler}}, \bibinfo {author} {\bibfnamefont
  {R.}~\bibnamefont {Lange}}, \bibinfo {author} {\bibfnamefont
  {N.}~\bibnamefont {Huntemann}}, \bibinfo {author} {\bibfnamefont
  {A.}~\bibnamefont {Surzhykov}}, \bibinfo {author} {\bibfnamefont {V.~A.}\
  \bibnamefont {Yerokhin}}, \bibinfo {author} {\bibfnamefont {J.~R.}\
  \bibnamefont {{Crespo L{\'o}pez-Urrutia}}}, \ and\ \bibinfo {author}
  {\bibfnamefont {P.~O.}\ \bibnamefont {Schmidt}},\ }\href@noop {} {\bibfield
  {journal} {\bibinfo  {journal} {Nature}\ }\textbf {\bibinfo {volume} {611}},\
  \bibinfo {pages} {43} (\bibinfo {year} {2022})}\BibitemShut {NoStop}%
\bibitem [{\citenamefont {Keller}\ \emph {et~al.}(2015)\citenamefont {Keller},
  \citenamefont {Partner}, \citenamefont {Burgermeister},\ and\ \citenamefont
  {Mehlst\"aubler}}]{Keller2015JAP}%
  \BibitemOpen
  \bibfield  {author} {\bibinfo {author} {\bibfnamefont {J.}~\bibnamefont
  {Keller}}, \bibinfo {author} {\bibfnamefont {H.~L.}\ \bibnamefont {Partner}},
  \bibinfo {author} {\bibfnamefont {T.}~\bibnamefont {Burgermeister}}, \ and\
  \bibinfo {author} {\bibfnamefont {T.~E.}\ \bibnamefont {Mehlst\"aubler}},\
  }\href@noop {} {\bibfield  {journal} {\bibinfo  {journal} {J. Appl. Phys.}\
  }\textbf {\bibinfo {volume} {118}},\ \bibinfo {pages} {104501} (\bibinfo
  {year} {2015})}\BibitemShut {NoStop}%
\bibitem [{\citenamefont {Keller}\ \emph {et~al.}(2019)\citenamefont {Keller},
  \citenamefont {Burgermeister}, \citenamefont {Kalincev}, \citenamefont
  {Didier}, \citenamefont {Kulosa}, \citenamefont {Nordmann}, \citenamefont
  {Kiethe},\ and\ \citenamefont {Mehlst\"aubler}}]{Keller2019PRA}%
  \BibitemOpen
  \bibfield  {author} {\bibinfo {author} {\bibfnamefont {J.}~\bibnamefont
  {Keller}}, \bibinfo {author} {\bibfnamefont {T.}~\bibnamefont
  {Burgermeister}}, \bibinfo {author} {\bibfnamefont {D.}~\bibnamefont
  {Kalincev}}, \bibinfo {author} {\bibfnamefont {A.}~\bibnamefont {Didier}},
  \bibinfo {author} {\bibfnamefont {A.~P.}\ \bibnamefont {Kulosa}}, \bibinfo
  {author} {\bibfnamefont {T.}~\bibnamefont {Nordmann}}, \bibinfo {author}
  {\bibfnamefont {J.}~\bibnamefont {Kiethe}}, \ and\ \bibinfo {author}
  {\bibfnamefont {T.~E.}\ \bibnamefont {Mehlst\"aubler}},\ }\href {\doibase
  10.1103/PhysRevA.99.013405} {\bibfield  {journal} {\bibinfo  {journal} {Phys.
  Rev. A}\ }\textbf {\bibinfo {volume} {99}},\ \bibinfo {pages} {013405}
  (\bibinfo {year} {2019})}\BibitemShut {NoStop}%
\bibitem [{\citenamefont {Chen}\ \emph {et~al.}(2017)\citenamefont {Chen},
  \citenamefont {Brewer}, \citenamefont {Chou}, \citenamefont {Wineland},
  \citenamefont {Leibrandt},\ and\ \citenamefont {Hume}}]{Chen2017PRL}%
  \BibitemOpen
  \bibfield  {author} {\bibinfo {author} {\bibfnamefont {J.-S.}\ \bibnamefont
  {Chen}}, \bibinfo {author} {\bibfnamefont {S.~M.}\ \bibnamefont {Brewer}},
  \bibinfo {author} {\bibfnamefont {C.~W.}\ \bibnamefont {Chou}}, \bibinfo
  {author} {\bibfnamefont {D.~J.}\ \bibnamefont {Wineland}}, \bibinfo {author}
  {\bibfnamefont {D.~R.}\ \bibnamefont {Leibrandt}}, \ and\ \bibinfo {author}
  {\bibfnamefont {D.~B.}\ \bibnamefont {Hume}},\ }\href {\doibase
  10.1103/PhysRevLett.118.053002} {\bibfield  {journal} {\bibinfo  {journal}
  {Phys. Rev. Lett.}\ }\textbf {\bibinfo {volume} {118}},\ \bibinfo {pages}
  {053002} (\bibinfo {year} {2017})}\BibitemShut {NoStop}%
\bibitem [{\citenamefont {Dub\'e}\ \emph {et~al.}(2013)\citenamefont {Dub\'e},
  \citenamefont {Madej}, \citenamefont {Zhou},\ and\ \citenamefont
  {Bernard}}]{Dube2013PRA}%
  \BibitemOpen
  \bibfield  {author} {\bibinfo {author} {\bibfnamefont {P.}~\bibnamefont
  {Dub\'e}}, \bibinfo {author} {\bibfnamefont {A.~A.}\ \bibnamefont {Madej}},
  \bibinfo {author} {\bibfnamefont {Z.}~\bibnamefont {Zhou}}, \ and\ \bibinfo
  {author} {\bibfnamefont {J.~E.}\ \bibnamefont {Bernard}},\ }\href@noop {}
  {\bibfield  {journal} {\bibinfo  {journal} {Phys. Rev. A}\ }\textbf {\bibinfo
  {volume} {87}},\ \bibinfo {pages} {023806} (\bibinfo {year}
  {2013})}\BibitemShut {NoStop}%
\bibitem [{\citenamefont {Dub\'e}\ \emph {et~al.}(2005)\citenamefont {Dub\'e},
  \citenamefont {Madej}, \citenamefont {Bernard}, \citenamefont {Marmet},
  \citenamefont {Boulanger},\ and\ \citenamefont {Cundy}}]{Dube2005PRL}%
  \BibitemOpen
  \bibfield  {author} {\bibinfo {author} {\bibfnamefont {P.}~\bibnamefont
  {Dub\'e}}, \bibinfo {author} {\bibfnamefont {A.~A.}\ \bibnamefont {Madej}},
  \bibinfo {author} {\bibfnamefont {J.~E.}\ \bibnamefont {Bernard}}, \bibinfo
  {author} {\bibfnamefont {L.}~\bibnamefont {Marmet}}, \bibinfo {author}
  {\bibfnamefont {J.-S.}\ \bibnamefont {Boulanger}}, \ and\ \bibinfo {author}
  {\bibfnamefont {S.}~\bibnamefont {Cundy}},\ }\href@noop {} {\bibfield
  {journal} {\bibinfo  {journal} {Phys. Rev. Lett.}\ }\textbf {\bibinfo
  {volume} {95}},\ \bibinfo {pages} {033001} (\bibinfo {year}
  {2005})}\BibitemShut {NoStop}%
\bibitem [{\citenamefont {Itano}\ \emph {et~al.}(1993)\citenamefont {Itano},
  \citenamefont {Bergquist}, \citenamefont {Bollinger}, \citenamefont
  {Gilligan}, \citenamefont {Heinzen}, \citenamefont {Moore}, \citenamefont
  {Raizen},\ and\ \citenamefont {Wineland}}]{Itano1993PRA}%
  \BibitemOpen
  \bibfield  {author} {\bibinfo {author} {\bibfnamefont {W.~M.}\ \bibnamefont
  {Itano}}, \bibinfo {author} {\bibfnamefont {J.~C.}\ \bibnamefont
  {Bergquist}}, \bibinfo {author} {\bibfnamefont {J.~J.}\ \bibnamefont
  {Bollinger}}, \bibinfo {author} {\bibfnamefont {J.~M.}\ \bibnamefont
  {Gilligan}}, \bibinfo {author} {\bibfnamefont {D.~J.}\ \bibnamefont
  {Heinzen}}, \bibinfo {author} {\bibfnamefont {F.~L.}\ \bibnamefont {Moore}},
  \bibinfo {author} {\bibfnamefont {M.~G.}\ \bibnamefont {Raizen}}, \ and\
  \bibinfo {author} {\bibfnamefont {D.~J.}\ \bibnamefont {Wineland}},\
  }\href@noop {} {\bibfield  {journal} {\bibinfo  {journal} {Phys. Rev. A}\
  }\textbf {\bibinfo {volume} {47}},\ \bibinfo {pages} {3554} (\bibinfo {year}
  {1993})}\BibitemShut {NoStop}%
\bibitem [{\citenamefont {Peik}\ \emph {et~al.}(2005)\citenamefont {Peik},
  \citenamefont {Schneider},\ and\ \citenamefont {Tamm}}]{Peik2005JPhysB}%
  \BibitemOpen
  \bibfield  {author} {\bibinfo {author} {\bibfnamefont {E.}~\bibnamefont
  {Peik}}, \bibinfo {author} {\bibfnamefont {T.}~\bibnamefont {Schneider}}, \
  and\ \bibinfo {author} {\bibfnamefont {C.}~\bibnamefont {Tamm}},\ }\href
  {\doibase 10.1088/0953-4075/39/1/012} {\bibfield  {journal} {\bibinfo
  {journal} {J. Phys. B}\ }\textbf {\bibinfo {volume} {39}},\ \bibinfo {pages}
  {145} (\bibinfo {year} {2005})}\BibitemShut {NoStop}%
\bibitem [{\citenamefont {Yu}\ \emph {et~al.}(2023)\citenamefont {Yu},
  \citenamefont {Sahoo},\ and\ \citenamefont {Suo}}]{YuSahSuo23}%
  \BibitemOpen
  \bibfield  {author} {\bibinfo {author} {\bibfnamefont {Y.-M.}\ \bibnamefont
  {Yu}}, \bibinfo {author} {\bibfnamefont {B.~K.}\ \bibnamefont {Sahoo}}, \
  and\ \bibinfo {author} {\bibfnamefont {B.-B.}\ \bibnamefont {Suo}},\
  }\href@noop {} {\bibfield  {journal} {\bibinfo  {journal} {Front. Phys.}\
  }\textbf {\bibinfo {volume} {11}},\ \bibinfo {pages} {1104848} (\bibinfo
  {year} {2023})}\BibitemShut {NoStop}%
\bibitem [{\citenamefont {VanArsdale}\ \emph {et~al.}(2024)\citenamefont
  {VanArsdale}, \citenamefont {Deutch}, \citenamefont {Lombardi}, \citenamefont
  {Nelson}, \citenamefont {Sherman}, \citenamefont {Spice}, \citenamefont
  {Yates}, \citenamefont {Yost},\ and\ \citenamefont
  {Brewer}}]{VanArsdale2024OptLett}%
  \BibitemOpen
  \bibfield  {author} {\bibinfo {author} {\bibfnamefont {J.~B.}\ \bibnamefont
  {VanArsdale}}, \bibinfo {author} {\bibfnamefont {M.~K.}\ \bibnamefont
  {Deutch}}, \bibinfo {author} {\bibfnamefont {M.}~\bibnamefont {Lombardi}},
  \bibinfo {author} {\bibfnamefont {G.}~\bibnamefont {Nelson}}, \bibinfo
  {author} {\bibfnamefont {J.}~\bibnamefont {Sherman}}, \bibinfo {author}
  {\bibfnamefont {J.}~\bibnamefont {Spice}}, \bibinfo {author} {\bibfnamefont
  {W.~C.}\ \bibnamefont {Yates}}, \bibinfo {author} {\bibfnamefont {D.~C.}\
  \bibnamefont {Yost}}, \ and\ \bibinfo {author} {\bibfnamefont {S.~M.}\
  \bibnamefont {Brewer}},\ }\href {\doibase 10.1364/OL.521175} {\bibfield
  {journal} {\bibinfo  {journal} {Opt. Lett.}\ }\textbf {\bibinfo {volume}
  {49}},\ \bibinfo {pages} {2545} (\bibinfo {year} {2024})}\BibitemShut
  {NoStop}%
\bibitem [{\citenamefont {Yudin}\ \emph {et~al.}(2014)\citenamefont {Yudin},
  \citenamefont {Taichenachev},\ and\ \citenamefont {Derevianko}}]{Yudin2014}%
  \BibitemOpen
  \bibfield  {author} {\bibinfo {author} {\bibfnamefont {V.~I.}\ \bibnamefont
  {Yudin}}, \bibinfo {author} {\bibfnamefont {A.~V.}\ \bibnamefont
  {Taichenachev}}, \ and\ \bibinfo {author} {\bibfnamefont {A.}~\bibnamefont
  {Derevianko}},\ }\href {\doibase 10.1103/PhysRevLett.113.233003} {\bibfield
  {journal} {\bibinfo  {journal} {Phys. Rev. Lett.}\ }\textbf {\bibinfo
  {volume} {113}},\ \bibinfo {pages} {233003} (\bibinfo {year}
  {2014})}\BibitemShut {NoStop}%
\bibitem [{\citenamefont {Tiedau}\ \emph {et~al.}(2024)\citenamefont {Tiedau},
  \citenamefont {Okhapkin}, \citenamefont {Zhang}, \citenamefont {Thielking},
  \citenamefont {Zitzer}, \citenamefont {Peik}, \citenamefont {Schaden},
  \citenamefont {Pronebner}, \citenamefont {Morawetz}, \citenamefont {De~Col},
  \citenamefont {Schneider}, \citenamefont {Leitner}, \citenamefont {Pressler},
  \citenamefont {Kazakov}, \citenamefont {Beeks}, \citenamefont {Sikorsky},\
  and\ \citenamefont {Schumm}}]{Tiedau2024}%
  \BibitemOpen
  \bibfield  {author} {\bibinfo {author} {\bibfnamefont {J.}~\bibnamefont
  {Tiedau}}, \bibinfo {author} {\bibfnamefont {M.~V.}\ \bibnamefont
  {Okhapkin}}, \bibinfo {author} {\bibfnamefont {K.}~\bibnamefont {Zhang}},
  \bibinfo {author} {\bibfnamefont {J.}~\bibnamefont {Thielking}}, \bibinfo
  {author} {\bibfnamefont {G.}~\bibnamefont {Zitzer}}, \bibinfo {author}
  {\bibfnamefont {E.}~\bibnamefont {Peik}}, \bibinfo {author} {\bibfnamefont
  {F.}~\bibnamefont {Schaden}}, \bibinfo {author} {\bibfnamefont
  {T.}~\bibnamefont {Pronebner}}, \bibinfo {author} {\bibfnamefont
  {I.}~\bibnamefont {Morawetz}}, \bibinfo {author} {\bibfnamefont {L.~T.}\
  \bibnamefont {De~Col}}, \bibinfo {author} {\bibfnamefont {F.}~\bibnamefont
  {Schneider}}, \bibinfo {author} {\bibfnamefont {A.}~\bibnamefont {Leitner}},
  \bibinfo {author} {\bibfnamefont {M.}~\bibnamefont {Pressler}}, \bibinfo
  {author} {\bibfnamefont {G.~A.}\ \bibnamefont {Kazakov}}, \bibinfo {author}
  {\bibfnamefont {K.}~\bibnamefont {Beeks}}, \bibinfo {author} {\bibfnamefont
  {T.}~\bibnamefont {Sikorsky}}, \ and\ \bibinfo {author} {\bibfnamefont
  {T.}~\bibnamefont {Schumm}},\ }\href {\doibase
  10.1103/PhysRevLett.132.182501} {\bibfield  {journal} {\bibinfo  {journal}
  {Phys. Rev. Lett.}\ }\textbf {\bibinfo {volume} {132}},\ \bibinfo {pages}
  {182501} (\bibinfo {year} {2024})}\BibitemShut {NoStop}%
\bibitem [{\citenamefont {Elwell}\ \emph {et~al.}(2024)\citenamefont {Elwell},
  \citenamefont {Schneider}, \citenamefont {Jeet}, \citenamefont {Terhune},
  \citenamefont {Morgan}, \citenamefont {Alexandrova}, \citenamefont
  {Tran~Tan}, \citenamefont {Derevianko},\ and\ \citenamefont
  {Hudson}}]{ElwellPRL2024}%
  \BibitemOpen
  \bibfield  {author} {\bibinfo {author} {\bibfnamefont {R.}~\bibnamefont
  {Elwell}}, \bibinfo {author} {\bibfnamefont {C.}~\bibnamefont {Schneider}},
  \bibinfo {author} {\bibfnamefont {J.}~\bibnamefont {Jeet}}, \bibinfo {author}
  {\bibfnamefont {J.~E.~S.}\ \bibnamefont {Terhune}}, \bibinfo {author}
  {\bibfnamefont {H.~W.~T.}\ \bibnamefont {Morgan}}, \bibinfo {author}
  {\bibfnamefont {A.~N.}\ \bibnamefont {Alexandrova}}, \bibinfo {author}
  {\bibfnamefont {H.~B.}\ \bibnamefont {Tran~Tan}}, \bibinfo {author}
  {\bibfnamefont {A.}~\bibnamefont {Derevianko}}, \ and\ \bibinfo {author}
  {\bibfnamefont {E.~R.}\ \bibnamefont {Hudson}},\ }\href {\doibase
  10.1103/PhysRevLett.133.013201} {\bibfield  {journal} {\bibinfo  {journal}
  {Phys. Rev. Lett.}\ }\textbf {\bibinfo {volume} {133}},\ \bibinfo {pages}
  {013201} (\bibinfo {year} {2024})}\BibitemShut {NoStop}%
\bibitem [{\citenamefont {Peik}\ and\ \citenamefont
  {Tamm}(2003)}]{Peik2003EPL}%
  \BibitemOpen
  \bibfield  {author} {\bibinfo {author} {\bibfnamefont {E.}~\bibnamefont
  {Peik}}\ and\ \bibinfo {author} {\bibfnamefont {C.}~\bibnamefont {Tamm}},\
  }\href {\doibase 10.1209/epl/i2003-00210-x} {\bibfield  {journal} {\bibinfo
  {journal} {Europhys. Lett.}\ }\textbf {\bibinfo {volume} {61}},\ \bibinfo
  {pages} {181} (\bibinfo {year} {2003})}\BibitemShut {NoStop}%
\bibitem [{\citenamefont {Bilous}\ \emph {et~al.}(2020)\citenamefont {Bilous},
  \citenamefont {Bekker}, \citenamefont {Berengut}, \citenamefont {Seiferle},
  \citenamefont {von~der Wense}, \citenamefont {Thirolf}, \citenamefont
  {Pfeifer}, \citenamefont {{Crespo L\'opez-Urrutia}},\ and\ \citenamefont
  {P\'alffy}}]{Bilous2020}%
  \BibitemOpen
  \bibfield  {author} {\bibinfo {author} {\bibfnamefont {P.~V.}\ \bibnamefont
  {Bilous}}, \bibinfo {author} {\bibfnamefont {H.}~\bibnamefont {Bekker}},
  \bibinfo {author} {\bibfnamefont {J.~C.}\ \bibnamefont {Berengut}}, \bibinfo
  {author} {\bibfnamefont {B.}~\bibnamefont {Seiferle}}, \bibinfo {author}
  {\bibfnamefont {L.}~\bibnamefont {von~der Wense}}, \bibinfo {author}
  {\bibfnamefont {P.~G.}\ \bibnamefont {Thirolf}}, \bibinfo {author}
  {\bibfnamefont {T.}~\bibnamefont {Pfeifer}}, \bibinfo {author} {\bibfnamefont
  {J.~R.}\ \bibnamefont {{Crespo L\'opez-Urrutia}}}, \ and\ \bibinfo {author}
  {\bibfnamefont {A.}~\bibnamefont {P\'alffy}},\ }\href {\doibase
  10.1103/PhysRevLett.124.192502} {\bibfield  {journal} {\bibinfo  {journal}
  {Phys. Rev. Lett.}\ }\textbf {\bibinfo {volume} {124}},\ \bibinfo {pages}
  {192502} (\bibinfo {year} {2020})}\BibitemShut {NoStop}%
\bibitem [{\citenamefont {Peik}\ \emph {et~al.}(2021)\citenamefont {Peik},
  \citenamefont {Schumm}, \citenamefont {Safronova}, \citenamefont {P\'alffy},
  \citenamefont {Weitenberg},\ and\ \citenamefont {Thirolf}}]{Peik2021}%
  \BibitemOpen
  \bibfield  {author} {\bibinfo {author} {\bibfnamefont {E.}~\bibnamefont
  {Peik}}, \bibinfo {author} {\bibfnamefont {T.}~\bibnamefont {Schumm}},
  \bibinfo {author} {\bibfnamefont {M.~S.}\ \bibnamefont {Safronova}}, \bibinfo
  {author} {\bibfnamefont {A.}~\bibnamefont {P\'alffy}}, \bibinfo {author}
  {\bibfnamefont {J.}~\bibnamefont {Weitenberg}}, \ and\ \bibinfo {author}
  {\bibfnamefont {P.~G.}\ \bibnamefont {Thirolf}},\ }\href {\doibase
  10.1088/2058-9565/abe9c2} {\bibfield  {journal} {\bibinfo  {journal} {Quantum
  Science and Technology}\ }\textbf {\bibinfo {volume} {6}},\ \bibinfo {pages}
  {034002} (\bibinfo {year} {2021})}\BibitemShut {NoStop}%
\end{thebibliography}

%

\end{document}